\documentclass[preprint,12pt]{aastex}
\usepackage{natbib}
\usepackage{epsfig}
\begin{document}

\title{A Compact Supermassive Binary Black Hole System}

\author{C. Rodriguez\altaffilmark{1,2}, {G. B. Taylor\altaffilmark{1,3}},
{R. T. Zavala\altaffilmark{4}}, {A. B. Peck\altaffilmark{5}},
{L. K. Pollack\altaffilmark{6}} \& {R. W. Romani\altaffilmark{7}}}

\altaffiltext{1}{Department of Physics and Astronomy, University of New Mexico, Albuquerque, NM 87131}
\altaffiltext{2}{Department of Physics, Universidad Simon Bolivar, Sartenejas, Venezuela}
\altaffiltext{3}{National Radio Astronomy Observatory, Socorro NM 87801}
\altaffiltext{4}{United States Naval Observatory, Flagstaff Station 10391 W. Naval Observatory Rd. Flagstaff, AZ 86001}
\altaffiltext{5}{Harvard-Smithsonian CfA, SMA Project, 645 N. A'ohoku Pl, Hilo, HI 96721}
\altaffiltext{6}{Department of Astronomy \& Astrophysics, University of California at Santa Cruz, Santa Cruz, CA 95064}
\altaffiltext{7}{Department of Physics, Stanford University, Stanford, CA 94305-4060}

\begin{abstract}

We report on the discovery of a supermassive binary black hole system
in the radio galaxy 0402+379, with a projected separation between the
two black holes of just 7.3 pc.  This is the closest black hole pair
yet found by more than two orders of magnitude.  These results are
based upon recent multi-frequency observations using the Very Long
Baseline Array (VLBA) which reveal two compact, variable,
flat-spectrum, active nuclei within the elliptical host galaxy of
0402+379. Multi-epoch observations from the VLBA also provide
constraints on the total mass and dynamics of the system.  Low spectral
resolution spectroscopy using the Hobby-Eberly Telescope indicates 
two velocity systems with a combined mass of the two
black holes of $\sim 1.5 \times 10^8\ {\rm M}_{\Sun}$.  The two
nuclei appear stationary while the jets emanating from the weaker of
the two nuclei appear to move out and terminate in bright hot spots.
The discovery of this system has implications for the number of
close binary black holes that might be sources of gravitational
radiation.  Green Bank Telescope observations at 22 GHz to search
for water masers in this interesting system are also presented.

\end{abstract}


\section{Introduction}
\label{intro}

Black holes are a direct consequence of the physics described in Einstein's
theory of gravity.
There is a great deal of indirect astronomical observational evidence
for these exotic objects in two mass ranges: stellar mass black holes,
with masses of 4-15 times the mass of our Sun; and supermassive black
holes, with masses ranging from $10^5$ to $10^{10}$ solar masses.
There is also some evidence for intermediate-mass black holes, those
with masses of a few hundred to a few thousand solar masses
(see Filippenko \& Ho 2003 \nocite{Filippenko03}, Gebhardt et al. 2005 \nocite{Gebhardt05}).

Since most nearby galaxies harbor supermassive black holes at their
centers \citep{Rich98}, the merging of galaxies, an essential part of the
galaxy formation process, is thought to be the prevalent method in
which supermassive black hole binaries are formed. Accordingly, such
systems should be common in galaxies. An understanding of the
evolution and formation of these systems is important for an
understanding of the evolution and formation of galaxies in general.

The evolution of a binary supermassive black hole involves three stages \citep{Begelman80}, 
which are summarized by \cite{Milo05} as follows: (1) As the
galaxies merge, the supermassive black holes sink toward the center of the new galaxy via dynamical 
friction forming a binary; (2) the binary continues to decay mainly due to
the interaction of stars on orbits intersecting the
binary, which are then ejected at velocities comparable to the binary's orbital velocity, 
carrying away energy and angular momentum; (3) finally, if the binary's separation decreases to
the point where the emission of gravitational waves becomes efficient at carrying
away the last remaining angular momentum, the supermassive black holes coalesce rapidly.
There is circumstantial
evidence that most binary black holes merge in less than a Hubble time
\citep{Komossa03}.  Therefore, the most massive systems that are able to coalesce
in less than a Hubble time will create the loudest
gravitational wave events in the universe \citep{Sesana04}, which
might be detectable by a low-frequency gravitational wave experiment
such as the Laser Interferometer Space Antenna (LISA).

Our ability to resolve both supermassive black holes in any given
binary system depends on the separation between them, and on their
distance from Earth. It is believed that the longest timescales in the
evolution of a supermassive binary black hole system leading up to
coalescence is the stage in which the system is closely bound
($\sim0.1 - 10$ pc), meaning that in most of these systems the black
hole pair can only be resolved by VLBI observations (see review by
Komossa 2003a \nocite{Komossa03} detailing observational evidence for supermassive
black holes binaries). Some source properties like X-shaped radio galaxies and
double-double radio galaxies, helical radio-jets,
double-horned emission line profiles, and semi-periodic variations
in lightcurves have been taken as indirect evidence
for compact binary black holes though other explanations are possible. 
The BL Lacertae Object  OJ 287 is a candidate for harboring a supermassive 
binary black hole, inferred from the characteristics of its optical lightcurve, 
which shows repeated outbursts at  11.86 y intervals \citep{Sillanpaa88}. Combining
optical as well as radio observations, \cite{Valtaoja00} presented a new interpretation
which suggests that at intervals of 11.86 y, the secondary black hole crosses
the accretion disk of the primary black hole, causing a thermal flare visible
only in the optical. About a year later, the disturbance propagates down 
the relativistic jet and results in the growth of new synchroton-emitting
shocks visible both in the optical and radio. The observed 11.86 y period
corresponds to the orbital period of the compact binary black hole. Some wider systems have, 
however, been
found more directly. The ultra luminous galaxy NGC 6240, discovered by the Chandra
X-ray observatory, was found to have a pair of active supermassive
black holes at its center \citep{Komo03}, separated by a
distance of 1.4 kpc. Another system that has been known for some time
is the double AGN (7 kpc separation) constituting the radio source 3C 75, which
was discovered by the VLA to have two pairs of radio jets
\citep{Owen85}.

In this paper we present further observations of the radio galaxy
0402+379, which was discovered by \cite{man03} to contain two central,
compact, flat spectrum, variable components (designated C1 and C2),
a feature which has not been observed in any other compact
source. \cite{man03} remarked upon the unusual properties found in
this source and proposed several physical explanations.  One
possible scenario is for one component to be a foreground or
background source, instead of being associated with 0402+379. However,
because of the small separation between C1 and C2 (7.3 pc), and
because of a faint bridge of radio emission found connecting
components C1 and C2, this theory was ruled out.  A second explanation
was that the nucleus was being gravitationally lensed. However, based
on the significant difference in the lightcurves of components C1 and
C2, and the close proximity of 0402+379 ($z = 0.055$), this theory was
also eliminated.  Two other scenarios were proposed by \cite{man03},
which could not be conclusively ruled out and remained as possible
explanations. The first of these suggest that component C2 could be a
knot in the southern jet, with C1 classified as the core. To test this
hypothesis we performed high frequency, high resolution Very Long Baseline
Array (VLBA\footnote {The National Radio Astronomy Observatory is operated by
Associated Universities, Inc., under cooperative agreement with the
National Science Foundation.})
observations of 0402+379, designed to resolve any jet component and to
look for relative motions.  The final explanation is that C1 and C2
are two active nuclei of a supermassive binary black hole system.

Throughout this discussion, we assume H$_{0}$=75 km s$^{-1}$
Mpc$^{-1}$, q$_{0}$ = 0.5, and 1 mas = 1.06 pc
    
\section{Observations}

\subsection{VLBA Observations from 2005}
\label{Observations}

VLBA observations were made on 2005 January 24 and June 13 at 0.317,
4.976, 8.410, 15.354, 22.222, and 43.206 GHz. Four IFs with a bandwidth
of 8 MHz were observed in 32 channels in both R and L circular
polarizations. Four-level quantization was employed at all six
frequencies.  The net integration time on 0402+379 was 115 minutes at
0.3 GHz, 69 minutes at 5 GHz, 69 minutes at 8 GHz, 122 minutes at 15
GHz, 251 minutes at 22 GHz, and 249 minutes at 43 GHz.  Standard
flagging, amplitude calibration, fringe fitting, bandpass calibration
(using 3C 84 for bandpass calibration and 3C 111 for gain
calibration), and frequency averaging procedures were followed in the
Astronomical Image Processing System (AIPS; van Moorsel et al. 1996).
Opacity corrections were performed for the 22 and 43 GHz data.
AIPS reduction scripts described in \citet{Ulvestad01} were used for a
large part of the reduction.  All manual editing, imaging,
deconvolution, and self-calibration were done using Difmap
\citep{Shep95}.

\subsection{Archival Observations}
\label{Archival}

To further study this source, we obtained fully-calibrated VLBI data
taken in 1990 \citep{Xu95}, in 1996 (VCS; Beasley et al. 2002), in
three epochs (1994, 1996, and 1999) of the CJ Proper Motion Survey
(Britzen et al. 2003) and in 2003 \citep{man03}. These data
were imaged and modeled in Difmap to aid in analysis of motions,
variability, and spectra of 0402+379. Further information regarding
these observations can be found in Table \ref{Observations}.

\subsection{Green Bank Telescope constraints on H$_2$O Masers}

Observations were made with the GBT on 2005 October 14.  We used the
18-22 GHz K-band receiver, which uses dual beams separated by
3$\arcmin$ in azimuth. The GBT beamwidth is $\sim36\arcsec$ at 22 GHz,
and pointing uncertainties were $\sim5\arcsec$. Pointing was corrected
hourly using 0402+379 itself, which has sufficient continuum emission.
The telescope was nodded between two positions on the sky such that
one beam was always centered on the position of 0402+379 during
integration. The spectrometer was configured with two bandpasses of
200 MHz each, overlapped by 20 MHz at the redshifted H$_2$O frequency
of 21.075 GHz, so that the total coverage was $\pm$ 2700 km s$^{-1}$
with respect to the systemic velocity of the galaxy. The spectral
resolution was equivalent to 0.33 km s$^{-1}$. The zenith system
temperature was between 40 and 55 K for the duration of the
run. Atmospheric opacity at 22 GHz was estimated from system
temperature and weather data, and ranged from 0.09 to 0.12 at the
zenith. The data were reduced using GBTIDL.  
We found that the higher frequency IF was subject to a 60 MHz ripple across the baseband, 
most likely the result of known imperfections and temperature sensitivities of the IF transmission.  
To reduce the effects of the bandpass ripple we subtracted polynomials of order 4 and 8 from the first 
and second IFs respectively.  We were left with some small residual ripples, but the period of 60 MHz 
is large enough that any maser emission present would have been apparent nonetheless.
We Hanning-smoothed the spectra following calibration.  The 1$\sigma$ rms sensitivity of these
observations is $\sim$2 mJy per km s$^{-1}$.  No maser emission was
detected.

\subsection {HET Spectroscopy}

We obtained a spectrum of the core of 0402+379 on 2004 December 11 with the 9.2m
Hobby-Eberly telescope (HET; Ramsey et al 1998\nocite{Ramsey98})
Marcario Low Resolution Spectrograph (LRS; Hill et
al. 1998\nocite{Hill98}). Two 600s exposures were taken, using the G3
VPH Grism, a Schott OG 515 blocking filter and a 1.5$^{\prime\prime}$
slit placed at the parallactic angle. The resulting spectrum covers
$\lambda\lambda = 6300-9120$ $\AA$ at $5.6$ $\AA$ resolution.  We applied
standard IRAF calibrations and find that the spectrum is similar to
that obtained by \cite{Stickel93}, with a reddened continuum and
Seyfert 2 emission lines at a redshift of $z=0.05523(1)$ (16,460
km s$^{-1}$).

\begin{figure}
\centering
\scalebox{0.75}{\includegraphics{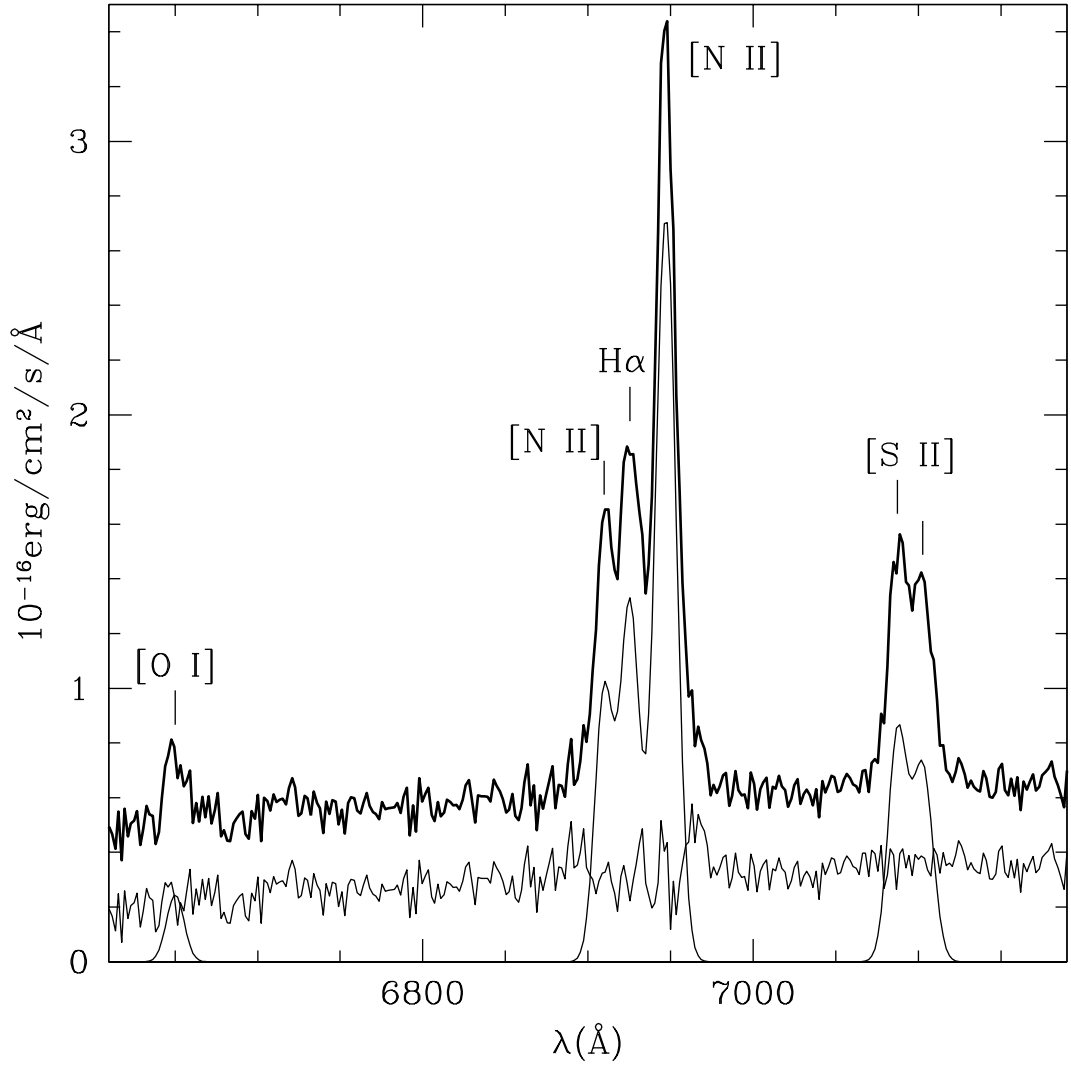}}
\vspace{1cm}
\caption{The optical spectrum at 5.6 $\AA$ spectral resolution taken by 
the Hobby-Eberly Telescope.  The thick line is the spectrum, the 
thin line shows the single component model and the residual spectrum.  }
\label{HET_Spectrum}
\end{figure}

In Figure \ref{HET_Spectrum} we show the H$\alpha$ region of the spectrum, 
where the only strong lines are present. The lines are resolved 
with Gaussian width $12.5\pm 1$ $\AA$, after deconvolution of the 
instrumental resolution. Here the uncertainty is the range in the fitted linewidths 
for the various species; this substantially exceeds the statistical
error. The lines appear asymmetric with a red shoulder suggesting two 
components with a $\sim 7\pm 1$ $\AA$ separation, i.e. a component
velocity separation of $\sim 300$ km s$^{-1}$. The residual to the line
fit in Figure \ref{HET_Spectrum} also shows significant excesses in the line wings,
suggesting the presence of a broader H$\alpha$ component.

\subsection{The Host Galaxy}
\label{HostGalaxy}

Optical imaging is at present quite limited, but in the Palomar Sky survey images 0402+379
appears as a relatively bright r = 17.2 elliptical galaxy embedded in a halo of patchy 
faint emission, extending to an apparent companion 25'' to the NE. Line emission 
(O{\sc i} \& N{\sc ii}) is seen from this region in the HET spectra, suggesting the presence of disturbed 
photo-ionized gas. An optical image in \cite{Stickel93} also shows this faint 
emission. These authors comment that the elliptical core has a ``flat brightness distribution''.  
The 2MASS \citep{2mass} J$-$K color of 0402+379 is 1.757, and this 
is consistent with the value expected for quasars at 0402+379's redshift \citep{Bark01}
The J$-$K color could indicate star formation activity from a recent merger, or an 
obscured AGN consistent with the radio galaxy identification for this source. Additional 
optical imaging would be useful to understand the dynamical state of 0402+379's core and 
the origin of the excitation of the surrounding nebulosity.

A source appears near 0402+379 in the ROSAT Bright source catalog
\citep{rosat}. 1RXSJ040547.3+380308 gives 0.21 PSPC cnts s$^{-1}$. Using
the Galactic absorption in this direction $N_H \approx 3 \times 10^{21}\ 
{\rm cm^{-2}}$, and assuming a typical AGN power law index of 
$\Gamma=1.7$ this corresponds to an unabsorbed 0.1-2 keV flux of $8.5 \times 10^{-12}\ 
{\rm erg\ cm^{-2}\ s^{-1}}$ or a luminosity $L_{0.1-2} \approx 5 \times 10^{43}\ 
{\rm erg\ s^{-1}}$, i.e. $\sim 2 \times 10^{-26}\ {\rm erg\ cm^{-2}\ Hz^{-1}}$. The 
radio flux density of 0402+379 at 5 GHz is 1.1 Jy \citep{pt,bwe}
which is a radio luminosity of $7\times10^{31}$ ergs sec$^{-1}$ Hz$^{-1}$.
This is in agreement with the X-ray to 5 GHz radio luminosity correlation
of \citet{brink}.

The ROSAT archive also contains two HRI 
pointings of this source for a combined exposure of 27 ks. These show that the 
X-ray source is largely resolved, extending over a radius of $\sim 15^{\prime\prime}$
and appears to follow the faint diffuse emission
surrounding the galaxy core. The bulk of the X-ray flux lies between 
the elliptical and its companion, further supporting the interaction 
hypothesis. The X-ray luminosity for this diffuse emission is comparable to the AGN
estimate above, e.g. $L_{0.1-2} \approx 3 \times 10^{43}\ {\rm erg\ s^{-1}}$
for a $1$ keV Raymond-Smith plasma, and likely dominates the flux of the 
ROSAT All-Sky Survey source.

We find that the X-ray emission of 0402+379 is unique; of the
35 known CSO's (Compact Symmetric Objects) this is the only source detected in the ROSAT All-Sky 
Survey. Thus, further X-ray observations could abet optical data in probing the nature of this 
emission and the connection with recent merger and/or nuclear activity.

\section{Results}

\subsection{Radio Continuum}
\label{RadioCont}

\begin{figure}
\centering
\scalebox{0.75}{\includegraphics{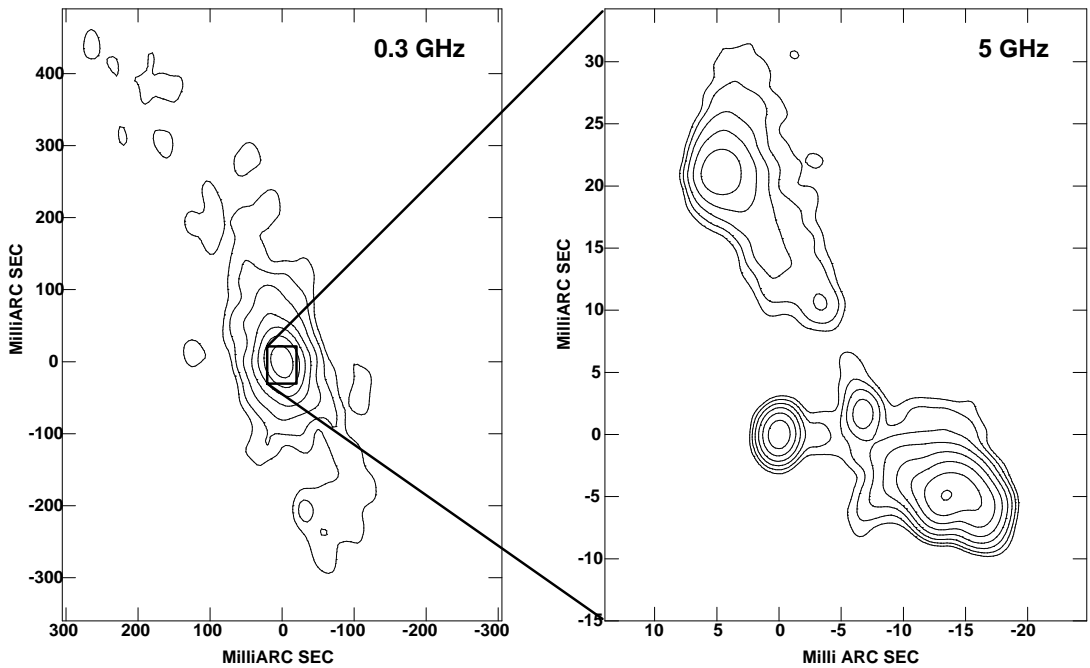}}
\vspace{-10cm}
\caption{Naturally weighted 2005 VLBA images
of 0402+379 at 0.3 and 5 GHz.  
Contours are drawn beginning at 3$\sigma$ and increase
by factors of 2 thereafter. 
The peak flux density and rms noise for each frequency
are given in Table \ref{Observations}. Right ascension and declination
are relative to C1, as shown in Figure \ref{VLBA05maps2}, assumed to be at 
$\sim$ RA 04h05m49.2623s Dec +38$^o$03$^{\prime}$32$^{\prime\prime}$.235 \citep{Beasley02}.}
\label{VLBA05maps1}
\end{figure}

   Figure \ref{VLBA05maps1} shows naturally weighted 0.3 and 5 GHz
images from the 2005 VLBA observations. The structure of the source at
5 GHz reveals the presence of two diametrically opposed jets, as well
as two central strong components, one directly between the jets and
the other one also between the jets but offset from the
center. Following the convention established by \cite{man03}, we
designated the aligned central component as C2, and the offset central
component as C1. As we can see in Figure \ref{VLBA05maps1}, the 5 GHz
image spans $\sim40$ mas ($\sim40$ pc), corresponding to a small
region in the central part of the 0.3 GHz image, which shows structure
on scales of $\sim500$ mas ($\sim500$ pc). The orientation of the 0.3
GHz image is consistent with that seen by the VLA at 5 GHz
\citep{man03}. The VLA image shows extended emission going northwards,
whereas the northern jet seen in our 5 GHz image is pointing in the
northeast direction, suggesting that at some point the jet changes
direction and starts moving northwards, probably as a consequence of
interactions with the surrounding medium.  The VLA 1.5 GHz image
\citep{man03} also shows the extended emission going northwards. The
southern jet seen in our 5 GHz image is pointing in the southwest
direction, which is consistent with both the 1.5 GHz and the 5 GHz VLA
images.

\begin{figure}
\centering
\scalebox{0.75}{\includegraphics{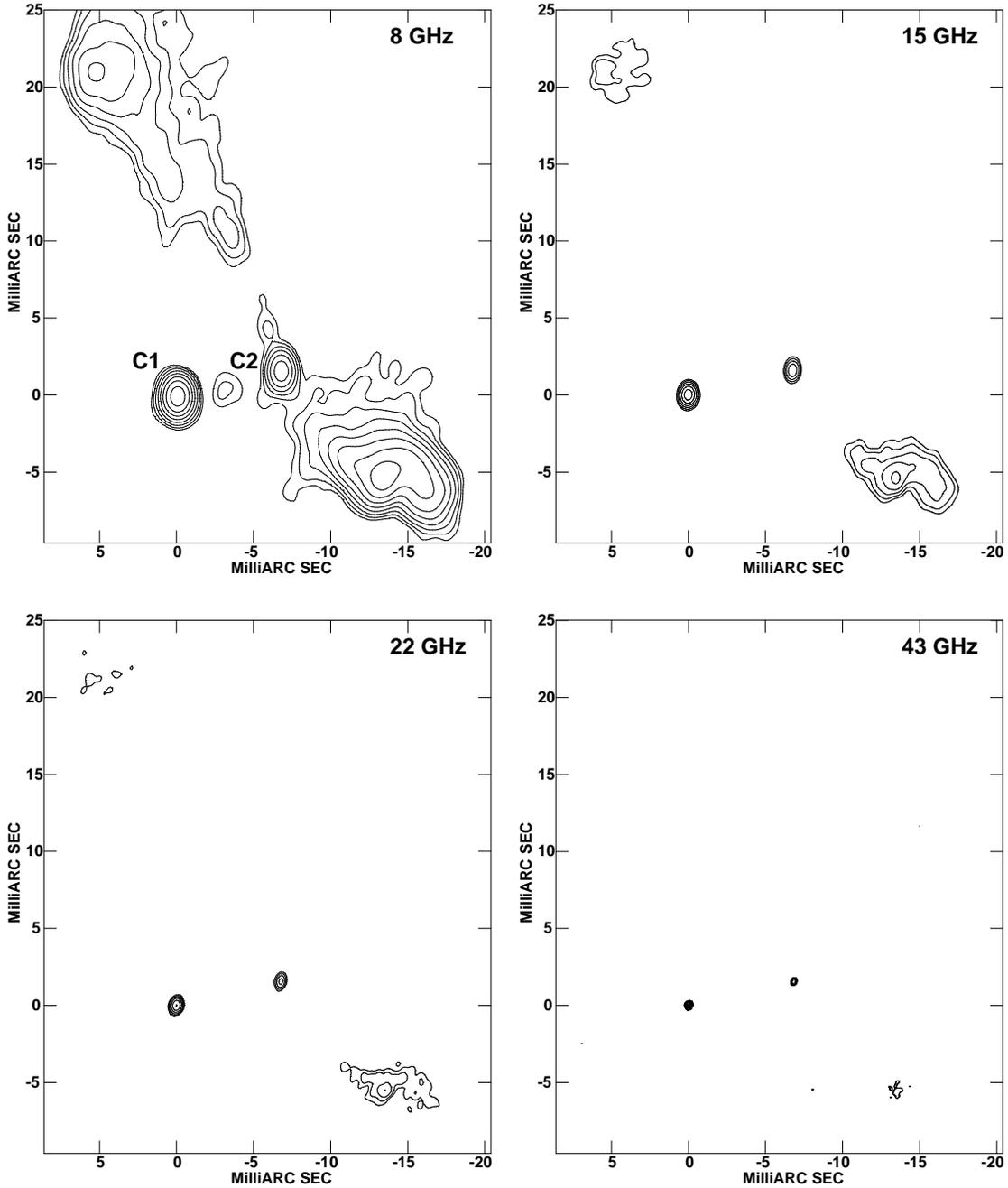}}
\vspace{-1.8cm}
\caption{Naturally weighted 2005 VLBA images 
of 0402+379 at 8, 15, 22 and 43 GHz. 
Contours are drawn beginning at 3$\sigma$ and increase
by factors of 2 thereafter.  
The peak flux density and rms noise for each frequency 
are given in Table \ref{Observations}.  
The labels shown in the 5 GHz map indicate the 
positions of the two strong, compact, central components derived from model-fitting.
}\label{VLBA05maps2}
\end{figure}

   Figure \ref{VLBA05maps2} shows naturally weighted 8, 15, 22, and 43 GHz images from the 2005
VLBA observations. For both the 8 and 15 GHz images, the overall structure of the source is
similar to that at 5 GHz, both jets are present as well as the two central components 
(C1 and C2). 

   It is clear that for higher frequencies the two central components
are easily distinguished and remain unresolved, while both jets become
fainter and are heavily resolved. At 22 GHz these effects are readily
apparent, and become more prominent at 43 GHz, where the jets can
barely be detected. Before these observations were made, the highest frequency
for which data had been taken for this source was 15 GHz.

  Elliptical or circular Gaussian components were fitted to the
visibility data using Difmap.  We obtained an estimate for the sizes
of C1 and C2 based on our 15, 22, and 43 GHz model fits. In this case,
we fit circular Gaussian components for both C1 and C2, obtaining an
average value of 0.173 $\pm$ 0.045 mas or 0.183 $\pm$ 0.048 pc for
C1, and 0.117 $\pm$ 0.033 mas or 0.124 $\pm$ 0.035 pc for C2.

\subsection{Component Motions and Variability}
\label{CompMotions&Var}

   In order to explore questions pertaining to motion and variability
in 0402+379, we obtained fully calibrated 5 GHz VLBI data taken in
1990 \citep{Xu95}, as well as 5 GHz VLBA data taken in three epochs
(1994, 1996, and 1999) of the CJ Proper Motion Survey \citep{bri03}
and in 2003 \citep{man03}. Combining these data with our 2005
observations at 5 GHz, we were able to probe motion and variability in
this source over a time baseline of 15 y.
   
\begin{figure}
\centering
\scalebox{0.5}{\includegraphics{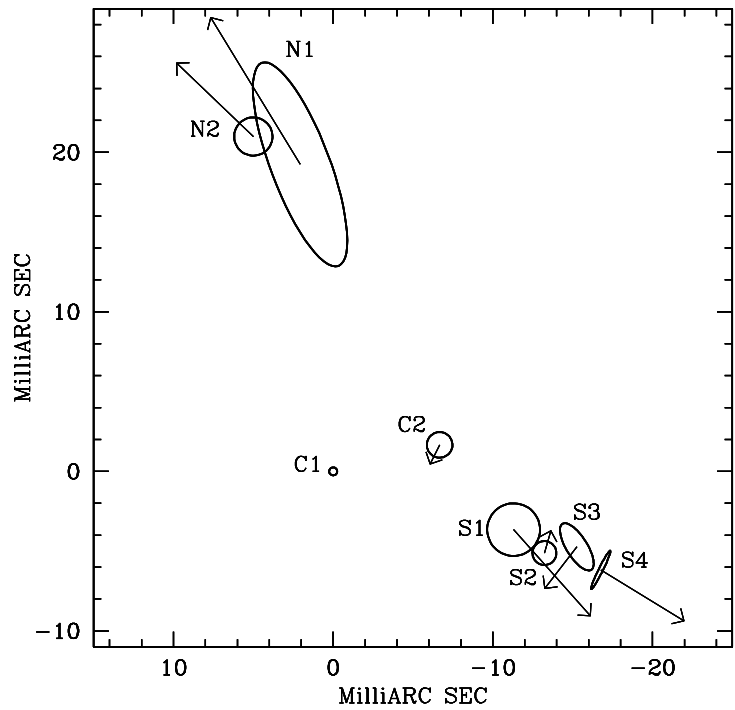}}
\vspace{-3cm}
\caption{Components model for the VLBA observations of 0402+379 at frequencies of 5 GHz and above. 
The arrows shown represent the direction of motion found for each component, relative to the 
position of C1, obtained from a 
time baseline of 15 y. Arrow lengths have been multiplied by a factor of 200.
See \S\ref{CompMotions&Var} for more details. Note that we placed 
arrows even for those components for which we are not claiming motion. The specific model shown corresponds
to a frequency of 5 GHz.  }
\label{VLBA05Model}
\end{figure}

  Motion and variability studies were performed by fitting eight
elliptical Gaussian components in Difmap to the 2003 visibility
data. Then, we used this model to fit the 5 GHz data corresponding to
the 1990, 1994, 1996, 1999, and 2005 epochs. We let only position and
flux density vary; all other parameters were held fixed at the 2003
values. Results from our fits are listed in Table \ref{Big_Gaussian},
and Figure \ref{VLBA05Model} shows the components model, where we have
labeled each of them and we also added arrows showing the direction of
motion of each component, which will be explained below.
  
\begin{figure}
\centering
\scalebox{0.5}{\includegraphics{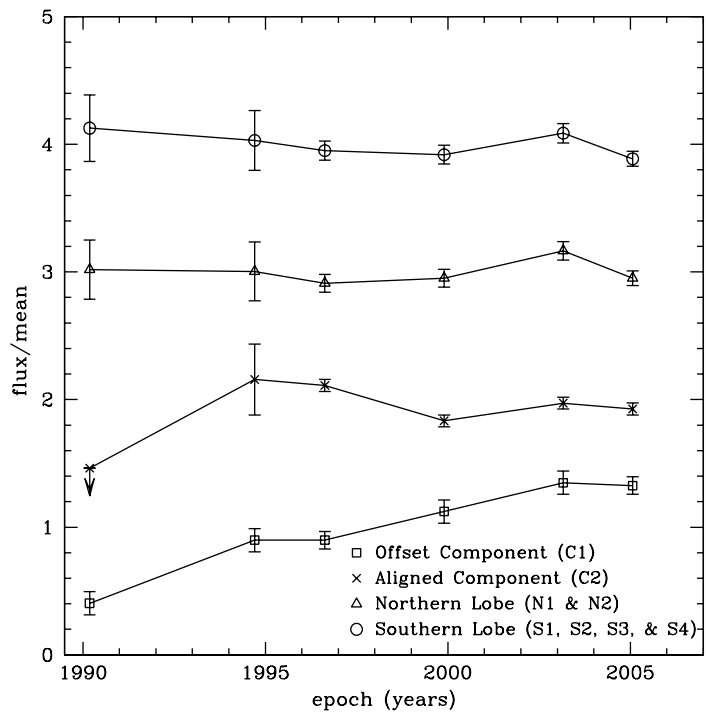}}
\vspace{-3cm}
\caption{Light curves of the different components of 0402+379 at 5 GHz.
The flux densities that produced this graph were taken from Table \ref{Big_Gaussian} and are discussed in \S\ref{CompMotions&Var}.
The displayed light curves were created by dividing each region's flux at each epoch by the mean region flux found
from averaging all observed epochs. The aligned core candidate, the northern lobe, and the southern lobe
are displayed on the $y$-axis by 1, 2, and 3 units, respectively. Errors are estimated from the rms noise and the absolute
flux calibration errors for each epoch.}
\label{LightCurves}
\end{figure}

To study component variability in 0402+379, we compared the flux
density for components C1 and C2, the mean flux density of the
southern components (S1, S2, S3, and S4), and the mean flux density of
the northern components (N1 and N2) over each of our 6 epochs at 5
GHz. The above regions were chosen primarily on the basis of their
isolation relative to other components in the source. Errors for each
region were computed on the basis of the rms noise and our estimated
absolute flux calibration errors ($\sim20\%$ for the 1990 and 1994 Mk
II VLBI epochs, and $\sim5\%$ for the 1996, 1999, 2003, and 2005 VLBA
epochs).  The resulting fractional variation light curves are shown in
Figure \ref{LightCurves}.  These light curves were created by dividing
each region's flux density at each epoch by the mean region flux
density found from averaging all epochs. To aid in readability of our
graph, the aligned component (C2), the northern lobe, and the southern
lobe are displaced on the $y$-axis by 1, 2, and 3 units, respectively.

From Figure \ref{LightCurves} and Table \ref{Big_Gaussian} we find
that component C1 substantially increases in flux over the 15 y
baseline, starting from 18 mJy in 1990 and increasing in brightness to
59 mJy in 2005.  We also find that component C2 is variable, ranging
from less than 10 mJy in 1990, to 24 mJy in 1996, and 20 mJy in
2005. Because our 1990 epoch was observed with Mk II VLBI, this
apparent variability could be in part attributed to poor data
quality. However, the measured flux for all other components in 1990
is quite consistent with that in our later epochs, suggesting that the
calculated upper limit for component C2's flux in 1990 is reliable and
that the observed variability in this component is significant. For
the southern and northern components, we find that there is no
substantial variation in the fluxes over the 15 y baseline.

Based on the time variability observed in both C1 and C2 we can
estimate the size of these components. However, in this case the
variability time scale found for C1 and C2 is quite long (roughly 5 to
10 y), which gives a weak upper limit on the sizes of the components
of a few parsecs, consistent with the estimate made in
\S\ref{RadioCont}.

\begin{figure}
\centering
\scalebox{0.5}{\includegraphics{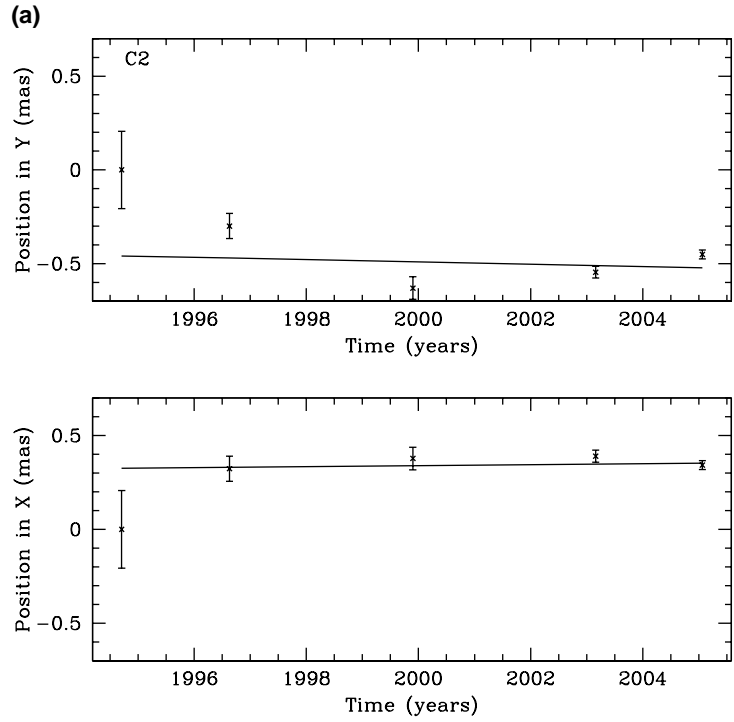}}
\vspace{10mm}
\caption{(a) Projected $x$ and $y$ positions of component C2 with time.  The $x$ and $y$ 
components of velocity are shown as a solid line. (b) Same for component N1. (c) Same for component N2.
(d) Same for component S1. (e) Same plot component S2. (f) Same component S3.
(g) Same for component S4. See Table \ref{Motions_Table} for more details.}
\label{C2}
\end{figure}

  To calculate the relative velocity of the components, we
chose component C1 as the reference, based on its strength,
compactness, and isolation relative to other components in the
source. We compared the relative motion for each epoch by fitting a
line to each component's relative position, split into $x$ and $y$
components, as a function of time. Results of this fitting process are
listed in Table \ref{Motions_Table}, and plotted in Figure
\ref{C2}.
  
The results of this analysis reveal significant motion for the
northern hot spots N1 and N2, yielding a value of 0.054 $\pm$ 0.008
mas/y, or (0.185 $\pm$ 0.008)c, and 0.033 $\pm$ 0.006 mas/y, or
(0.114 $\pm$ 0.019)c respectively. These results show that the
northern jet is moving away from the two central components to the
northeast.
 
   For the southern components S2 and S3, significant motion was also
found, yielding 0.0073 $\pm$ 0.0025 mas/y, or (0.0251 $\pm$
0.0085)c, and 0.016 $\pm$ 0.003 mas/y, or (0.056 $\pm$ 0.010)c
respectively. For the other two southern components, S1 and S4, even
though the values found for the velocities were larger than those
found for S2 and S3, the values obtained in the fitting for $\chi^{2}$
were large, a fact that can be verified by looking at Figures \ref{S1}(d) and \ref{S1}(g). 
However, for the projected $x$ position of component S4
with time, the $\chi^{2}$ obtained was nearly unity, which gives us
some confidence in the motion of this component, at least in the $x$
direction. We conclude then that, on average, the southern jet is
moving away from the two central components to the southwest, though
more slowly than the northern jet.
 
  The results obtained for C2 show no significant motion. The value
obtained for the limit on the motion of this component is equal to
0.0067 $\pm$ 0.0094 mas/y, or less than 0.088c.

   In Figure \ref{VLBA05Model} we draw arrows showing the direction of
motion found for each component, as well as their relative
magnitude. It is important to note that we placed arrows even for
those components for which we do not claim significant motion.

\subsection{Radio Continuum Spectra}
\label{RadioContSpec}

  By appropriately tapering our 22 GHz 2005 data, we obtained an image
resolution matched to our 8 GHz continuum image. These two images were
then combined to generate an image of the spectral index distribution
across the source (Figure \ref{VLBA05spix8-22}).  In both hotspots of
the source, N2 and S2, a steep spectrum was found, whereas in both
central components, the spectrum is flat. A plot of flux density as a
function of frequency is included in the image of the spectral index
distribution for both C1 and C2. The values for the flux densities
used in order to make these plots were measured from matching
resolution images.  Details regarding these results are listed in
Table \ref{Peak_Fluxes}.

\section{Discussion}
\label{Discussion}

 Four possible scenarios were proposed by \cite{man03} in order to
explain the unusual properties found in 0402+379.  Two of them were
ruled out, but the possibility that C1 or C2 was an unusual jet
component in a dense ISM could not be conclusively eliminated. 

Our high resolution
observations confirm the compactness of component C2, and measure a
size of 0.124 $\pm$ 0.035 pc. C2 is found to have no significant
motion, whereas significant flux density variability is found.
The spectral peak is shown to be at $\sim10$ GHz.  It is possible
that either or both C1 and/or C2 could be a jet component lit up
in a collision with a dense interstellar medium.  In this scenario
the low observed velocity ($<0.088$c) is due to the impact with
the ISM, and the spectrum is modified by local acceleration of
particles.  Difficulties with the jet component explanation are
(1) it requires a dramatic change in the jet axis on timescales
of a few 10s of years, while the larger scale emission (see Figure \ref{VLBA05maps1})
indicates that the jet axis has been fairly stable on time
scales up to 10$^{4}$ y; (2) a large gradient in density
needs to be invoked to decelerate C1 and/or C2 but allow the
hotspots to advance \citep{man03}; (3) if C2 is the core responsible
for the observed jets and hot spots, and C1 is a jet component, then
the counterjet is conspicuous in its absence given the orientation of
the source close to the plane of the sky indicated by the source symmetry.

The absence of a jet associated with C1 might at first
seem unusual since only 6 of 87 (7\%) sources in the First Caltech-Jodrell Bank
VLBI  survey (CJ1 - the survey in which 0402+379 was imaged by Xu et al. 1995
\nocite{Xu95}) show naked cores with no sign of a jet at 5 GHz.
At lower flux
levels, however, which are more appropriate since C1 by itself would not have made
it into the CJ1 sample, the fraction of naked cores increases to 6 of 24 sources
(25\% - Taylor et al. 2005 \nocite{Taylor05}).  Thus the absence of a jet from C1 is not
by itself evidence against the identification of C1 as an AGN.

The characteristics found in C1 and C2 are typical of AGN; in a
complete survey of 32 sources imaged at 43 GHz, \citep{lis01} found no
unresolved, isolated jet components.  This, together with the
morphology of 0402+379, leads us to surmise that neither C1 nor C2 is
likely to be a jet component.  This leaves the remaining most likely
explanation that C1 and C2 are both active black holes in a compact
system.  For the remainder of the discussion we assume that this is
the case and explore what we can learn about such a system from the
present observations.

\subsection{Constraints on the mass of the black holes}
\label{MassBH}

We can use the HET and the VLBA observations to obtain kinematic constraints
on the mass of the black hole system. At our observed separation, an 
orbital velocity of 300\,km/s implies a system mass of 
$ 1.5 \times 10^8$($v/300$ km s$^{-1}$)$^2$ ($r/7.3$ pc) M$_{\Sun}.$
The suggested line splitting from the optical spectra thus implies a 
mass of a few times $10^8\ {\rm M}_{\Sun}$. If both radio nuclei are also optically
active and show comparable emission line strengths, as expected, then 
the observed Gaussian line widths put a limit on the line FWHM of $\sim 
1300$ km s$^{-1}$, implying a limit on the combined black hole mass of $\sim 3 \times 
10^{9}\ {\rm M}_{\Sun}$. Note that the H{\sc i} absorption profiles of \cite{man03} also
found velocity structure of $\sim 1000\ $km s$^{-1}$ extending over a transverse 
distance of $\sim 20$ pc. 
On this scale the mass contribution from gas and stars is not
significant, so the limit obtained reflects the mass of the binary black
hole system and indicates a high mass system.
The implied mass, $\sim5\times10^9\ {\rm M}_{\Sun}$, could 
be dominated by the AGN, if the absorption occurs in the nuclear 
region. Alternatively the two velocity systems might be probing line-of-sight
velocities in the products of a recent merger.

An alternative estimate for the central compact object mass can
be derived from the blue luminosity of the host bulge \citep{Kormendy01}. 
The observed V magnitude of 0402+379 is $\sim 18.5$ \citep{Wills73}, 
after correcting for the Galactic ${\rm A_V}\sim 1.6$ and
a median B-V color index of $\sim 0.9$ for 484 ellipticals in the Uppsala 
General Catalogue (UGC), we obtain a B magnitude of 17.8. 
This corresponds to a central compact object 
mass of $\sim 7 \times 10^7\ {\rm M}_{\Sun}$, in reasonable accord with the estimates
above, given that this may still be a disturbed system.

\subsection{Supermassive Binary Black Hole Orbital Parameters}
\label{smbh}

Using our system mass estimate equal to 1.5 $\times 10^8$ M$_{\Sun}$ and
the projected radial separation between them derived from our
2005 maps (7.3 pc), we find from Kepler's Laws that the period of
rotation for such a binary supermassive black hole system
should be $\sim1.5\times10^5$ y. This period corresponds
to a relative projected velocity between components C1 and C2 of
$\sim0.001$ c. The upper limit found for component C1, $<0.088$c, is
consistent with the expected relative velocity between C1 and C2,
assuming a stable, Keplerian orbit.  To actually constrain the masses
of the black holes would require observations
over a longer time baseline ($\sim100$ y). 

\subsection{Gravitational Wave Signal}
\label{GravWave}

What sort of gravitational wave signal might a binary supermassive
black hole in 0402+379 generate? Assuming the current separation is
7.3 pc and the total mass is $ 1.5 \times 10^8\ {\rm M}_{\Sun}$\ the natural
gravitational wave frequency \citep{hughes} is approximately
$2 \times 10^{-13}$ Hz. This is well below the expected minimum
frequency of LISA. Although 0402+379 may be a long way from generating
a detectable gravitational wave signal it may represent a source of
noise important for future observations of cosmologically produced
gravitational radiation. Ultra low frequency gravitational radiation
generated during inflation \citep{hughes} has an upper limit of
10$^{-13}$ Hz.  Thus, a population of black hole binaries like
0402+379 may generate substantial noise which could interfere with the detection of
the physics of inflation. An estimate of this noise contribution requires a population
synthesis model (e.g., Sesana et al. 2005) which is beyond
the scope of this work.

The final stage in the evolution of a
binary black hole system is the gravitational radiation stage, where
the semimajor axis decreases to the point at which gravitational
radiation becomes the dominant dissipative force. A binary black hole
on a circular orbit will merge within the time \citep{Peters64}:

\begin{equation}
\label{merge_time}
t_{\rm merge} (a) = 5.8\times10^6\left(\frac{a}{0.01{\rm pc}}\right)^4\left(\frac{10^8{\rm M}_{\Sun}}{m_1}\right)^3\frac{m_{1}^2}{m_2(m_1+m_2)} {\rm y}, 
\end{equation}

\noindent
where $m_1$ and $m_2$ are the masses of the black holes, and $a$ is
the separation between them, once the system is in the final stage.
Using the projected radial separation between the black holes 
we measure for 0402+379 (7.3 pc) and assuming they both have a mass equal to
$\sim10^8\ {\rm M}_{\Sun}$, we obtain a merger time equal to $ \sim 10^{18}
$ y. Some other loss of angular momentum will be necessary if this system
is to merge in less than a Hubble time.

It is not clear whether, at present, dynamical friction losses or gas 
dissipative effects (see Komossa 2003a \nocite{Komossa03}, 
Merritt \& Milosavljevi{\'c} 2005 \nocite{Milo05}) 
are strong enough to bring the binary to the gravitational radiation loss regime within a
Hubble time. This relates to the issue of the probability of catching 
the binary at its present (modest) separation. Are we seeing a recent 
merger in the act of core coalescence or has the binary stalled and 
must now await loss-cone replenishment and/or re-supply of nuclear gas? 
The fact that both nuclei are active (radio bright) suggests on-going 
accretion and implies dissipation today. Only a larger sample of 
imaged active nuclei can address the fraction of the population in this
state, for example the VLBA Imaging and Polarization
Survey (VIPS, Taylor et al. 2005 \nocite{Taylor05}) which will image 1169 sources.

\subsection{Jet Components}
\label{JetComp}

Assuming that C2 is the origin of the radio emission on parsec scales,
we can constrain the orientation of 0402+379.  In the simple beaming
model for simultaneously ejected jet components moving in opposite
directions, both the arm length ratio $ D $ and the flux density ratio
$ R $ depend on the intrinsic speed $\beta=v/{\rm c}$ and the angle of the
twin jets to the line of sight $\theta$
\citep{Taylor&Vermeulen97}. The arm length ratio, $D$, is given by

\begin{equation}
\label{arm_ratio}
D=\frac{\mu_{\rm N}}{\mu_{\rm S}}=\frac{d_{\rm N}}{d_{\rm S}}=\left( \frac{1+\beta \cos \theta}{1-\beta \cos \theta}\right),
\end{equation}

\noindent
where the apparent projected distances from C2 (assumed to be the origin of radio emission)
are $ d_{\rm N} $ for the northern jet (approaching side) and $ d_{\rm S} $ for the southern jet (receding side),
and the apparent proper motions are $ \mu_{\rm N} $ for the northern jet and $ \mu_{\rm S} $ for the southern jet.
Similarly, the flux density ratio, $R$, between the northern and southern jet is

\begin{equation}
\label{flux_ratio}
R=\frac{S_{\rm N}}{S_{\rm S}}=\left( \frac{1+\beta \cos\theta}{1-\beta \cos\theta}\right)^{k-\alpha},
\end{equation}

\noindent
where $ \alpha $ is the spectral index, $ k = 2 $ for a continuous jet, and $ k = 3 $ for discrete
jet components (c.f. Lind \& Blandford 1985\nocite{Lind&Blandford85}).

In order to determine the arm length ratio we calculated the distances
from N2 and S2 to C2 (using the values shown in Table
\ref{Big_Gaussian}), yielding a value of $ D = d_{\rm N} / d_{\rm S} = 2.42 $.

For the proper motion ratio we proceeded as follows. We determined a
flux density weighted average velocity, for both the northern and
southern components, with respect to C2, using the values shown in
Table \ref{Motions_Table} (calculating the $x$ and $y$ components
separately and then calculating the resultant value).  Following this
procedure we obtained a value of $ \mu_{\rm N} / \mu_{\rm S} = 6.12 $.

The flux density ratio was obtained dividing the sum of the fluxes of
the northern components by the sum of the fluxes of the southern
components (see Table \ref{Big_Gaussian}), yielding a value of $ S_{\rm N}
/ S_{\rm S} = 0.32 $.

If Doppler boosting dominates the appearance of 0402+379 then all the
ratios calculated should be in agreement according to Equations
\ref{arm_ratio} and \ref{flux_ratio}. However, using $ k =2 $ and
estimating a value for the spectral index of both the north and south
hotspots of $ \alpha=-1 $ (see Table \ref{Peak_Fluxes}), we find that
$ D = 2.42 $, $ \mu_{\rm N} / \mu_{\rm S} = 6.12 $, and $ R^\frac{1}{k-\alpha}
= 0.68 $, results that are far from agreement, thus ruling out
Doppler boosting as the dominant effect.  A possible explanation
for this discrepancy could be an enhancement in the density of the 
interstellar medium to the
south of the core, which will reduce the velocity in that direction; slow the
expansion over time, thus changing the arm length ratio; and
increase the flux density of the southern jet. Additional supporting evidence
for this hypothesis comes from the enhancement in H{\sc i} opacity to the
southwest \citep{man03}.  Assuming that the
density enhancement is a recent phenomenon, we can use the arm length
ratio obtained in addition to Equation \ref{arm_ratio} to get $ \beta
\cos \theta = 0.4 $. This result implies that the intrinsic velocity
must be at least 0.4c and $ \theta $ must be less than $66^{\circ}$.

Using the relative velocity between the northern and southern jets
(calculated from the values obtained for the flux density weighted
averaged velocities of both the southern and northern components) as
well as the distance between components S2 and N2, we estimated the
age of the current radio emission to be  $ \sim 500 $
y. If the slower jet velocities for the southern jet are used 
then the age for the southern lobe is three times as large. Note 
that these jet ages are much less than the $\sim 10^5$ y binary period
in \S\ref{smbh}, so we expect no significant orbital displacement of 
the jet components.

\section{Conclusion}

Based on the compactness, motion, variability, and spectra of the 
two central components in 0402+379 we conclude that they are
both active nuclei of a single galaxy.  This pair of AGN forms
the closest binary black hole system yet discovered with a
projected separation of 7.3 pc. The total mass of the system is
estimated to be $1.5 \times 10^8$ M$_\Sun$, and the gravitational radiation frequency
to be 2$\times10^{-13}$ Hz.  Energy losses due to gravitational
radiation are not yet significant, so that other mechanisms 
must be invoked if the orbit is to decay. 
0402+379 may be the tip of an iceberg for a population of 
supermassive black hole binaries with parsec scale separations. 
Such a population may produce significant gravitational wave radiation 
which may need to be considered for the detection of 
gravitational radiation in the ultra to very low frequency 
bands.

Having found one system in
the CJF sample \citep{Taylor96} of 293 sources, we might expect to
find others in a larger survey. The VIPS \citep{Taylor05} will image 1169 sources and hopefully
find additional compact binary black holes.

{\it Facilities:} \facility{VLBA}, \facility{GBT}, \facility{HET}

\acknowledgments 
We thank an anonymous referee for constructive comments. 
This research has made use of the
NASA/IPAC Extragalactic Database (NED) which is operated by the Jet
Propulsion Laboratory, Caltech, under contract with NASA.  The
National Radio Astronomy Observatory is a facility of the National
Science Foundation operated under a cooperative agreement by
Associated Universities, Inc.

\clearpage

\begin{deluxetable}{lccccccccc}
\tabletypesize{\scriptsize}
\tablecolumns{10}
\tablewidth{0pt}
\tablecaption{Observations\label{Observations}}
\tablehead{\colhead{Frequency}&\colhead{Instrument}&\colhead{Date}&\colhead{Time}&\colhead{BW}&\colhead{Pol.}&\colhead{IFs}&\colhead{Peak} &\colhead{rms} &\colhead{Reference}\\
\colhead{(GHz)}&\colhead{}&\colhead{}&\colhead{(min)}&\colhead{(MHz)}&\colhead{}&\colhead{}&\colhead{(Jy/beam)}&\colhead{(mJy)} &\colhead{}}
\startdata

0.32 & VLBA & 2005 Jun 13 & 115 & 8 & 2 & 4 & 0.32 & 0.59 & This paper \\
4.99 & VLBI Mk 2 & 1990 Mar 10 & 80 & 2 & 1 & 1 &0.17 &0.94 & Xu et al. 1995 \\
4.99 & VLBI Mk 2 & 1994 Sep 17 & 57 & 2 & 1 & 1 &0.17 &1.6 & Britzen et al. 2003 \\
4.99 & VLBI & 1996 Aug 19 & 41 & 8 & 1 & 1 &0.16 &0.63 &Britzen et al. 2003 \\
4.99 & VLBA & 1999 Nov 26 & 35 & 8 & 2 & 2 &0.18 &0.42 &Britzen et al. 2003\\
5.00 & VLBA & 2003 Mar 02 & 478 & 16 & 2 & 1 &0.16 &0.21 &Maness et al. 2003 \\
4.98 & VLBA & 2005 Jan 24 & 69 & 8 & 2 & 4 &0.13 &0.12 &This paper \\
8.41 & VLBA & 2005 Jun 13 & 69 & 8 & 2 & 4 &0.06 &0.06 &This paper \\
15.35 & VLBA &2005 Jan 24 & 122 & 8 & 2 & 4 &0.05 &0.17 &This paper \\
22.22 & VLBA &2005 Jun 13 & 251 & 8 & 2 & 4 &0.03 &0.22 &This paper \\
43.21 & VLBA &2005 Jan 24 & 249 & 8 & 2 & 4 &0.02 &0.25 &This paper \\
\enddata
\end{deluxetable}

\begin{deluxetable}{lcccccccc}
\tabletypesize{\scriptsize}
\tablecolumns{9}
\tablewidth{0pt}
\tablecaption{Gaussian Model Components\tablenotemark{*}.\label{Big_Gaussian}}
\tablehead{\colhead{Component}&\colhead{Epoch}&\colhead{$S$}&\colhead{$r$}
&\colhead{$\theta$}&\colhead{$a$}&\colhead{$b/a$}&\colhead{$\Phi$}
&\colhead{$\chi^{2}$} \\
\colhead{} & \colhead{} & \colhead{(Jy)} & \colhead{(mas)} 
& \colhead{($^o$)}&\colhead{(mas)}&\colhead{}&\colhead{($^o$)}}
\startdata
C1... & 1990.185 & 0.018 $\pm$ 0.004 &  0.0   &     0.0 & 0.525 & 1.00 &  172.4  & 1.13 \\
      & 1994.708 & 0.040 $\pm$ 0.009 &  0.0   &     0.0 & 0.525 & 1.00 &  172.4  & 0.90 \\
      & 1996.630 & 0.040 $\pm$ 0.003 &  0.0   &     0.0 & 0.525 & 1.00 &  172.4  & 1.31 \\  
      & 1999.899 & 0.050 $\pm$ 0.004 &  0.0   &     0.0 & 0.525 & 1.00 &  172.4  & 1.28 \\
      & 2003.162 & 0.060 $\pm$ 0.004 &  0.0   &     0.0 & 0.525 & 1.00 &  172.4  & 2.07 \\
      & 2005.062 & 0.059 $\pm$ 0.003 &  0.0   &     0.0 & 0.525 & 1.00 &  172.4  & 1.67 \\
C2... & 1990.185 &    $<$0.010        &         &       &      &         &      \\
      & 1994.708 & 0.025 $\pm$ 0.006 &  7.323 &  286.21 & 1.61  & 1.00 &  207.9 & 0.90 \\
      & 1996.630 & 0.024 $\pm$ 0.002 &  6.932 &  284.58 & 1.61  & 1.00 &  207.9 & 1.31 \\
      & 1999.899 & 0.018 $\pm$ 0.001 &  6.803 &  282.00 & 1.61  & 1.00 &  207.9 & 1.28 \\ 
      & 2003.162 & 0.021 $\pm$ 0.001 &  6.809 &  282.72 & 1.61  & 1.00 &  207.9 & 2.07 \\
      & 2005.062 & 0.020 $\pm$ 0.001 &  6.876 &  283.40 & 1.61  & 1.00 &  207.9  & 1.67 \\
S1... & 1990.185 & 0.078 $\pm$ 0.017 & 11.063 &  254.74 & 3.29  & 1.00 &  308.3  & 1.13 \\
      & 1994.708 & 0.090 $\pm$ 0.020 & 12.154 &  251.22 & 3.29  & 1.00 &  308.3  & 0.90 \\
      & 1996.630 & 0.087 $\pm$ 0.006 & 11.978 &  251.46 & 3.29  & 1.00 &  308.3  & 1.31 \\
      & 1999.899 & 0.090 $\pm$ 0.006 & 11.600 &  251.83 & 3.29  & 1.00 &  308.3  & 1.28 \\
      & 2003.162 & 0.119 $\pm$ 0.007 & 11.846 &  251.19 & 3.29  & 1.00 &  308.3  & 2.07 \\
      & 2005.062 & 0.103 $\pm$ 0.005 & 11.876 &  252.10 & 3.29  & 1.00 &  308.3  & 1.67 \\
S2... & 1990.185 & 0.291 $\pm$ 0.064 & 14.119 &  248.45 & 1.51  & 1.00 &  302.3  & 1.13 \\
      & 1994.708 & 0.228 $\pm$ 0.050 & 14.364 &  248.46 & 1.51  & 1.00 &  302.3  & 0.90 \\
      & 1996.630 & 0.211 $\pm$ 0.015 & 14.234 &  248.71 & 1.51  & 1.00 &  302.3  & 1.31 \\ 
      & 1999.899 & 0.198 $\pm$ 0.014 & 14.120 &  248.66 & 1.51  & 1.00 &  302.3  & 1.28 \\
      & 2003.162 & 0.208 $\pm$ 0.012 & 14.165 &  248.88 & 1.51  & 1.00 &  302.3  & 2.07 \\
      & 2005.062 & 0.159 $\pm$ 0.008 & 14.192 &  248.84 & 1.51  & 1.00 &  302.3  & 1.67 \\
S3... & 1990.185 & 0.187 $\pm$ 0.041 & 16.029 &  253.40 & 3.40  & 0.37 &   32.0  & 1.13 \\
      & 1994.708 & 0.168 $\pm$ 0.037 & 16.144 &  252.75 & 3.40  & 0.37 &   32.0  & 0.90 \\ 
      & 1996.630 & 0.158 $\pm$ 0.011 & 16.149 &  252.48 & 3.40  & 0.37 &   32.0  & 1.31 \\
      & 1999.899 & 0.152 $\pm$ 0.011 & 16.084 &  252.39 & 3.40  & 0.37 &   32.0  & 1.28 \\
      & 2003.162 & 0.186 $\pm$ 0.011 & 16.015 &  252.48 & 3.40  & 0.37 &   32.0  & 2.07 \\
      & 2005.062 & 0.153 $\pm$ 0.008 & 15.980 &  252.48 & 3.40  & 0.37 &   32.0  & 1.67 \\
S4... & 1990.185 & 0.006 $\pm$ 0.001 & 17.454 &  251.27 & 2.72  & 0.10 &  333.8  & 1.13 \\
      & 1994.708 & 0.028 $\pm$ 0.006 & 17.399 &  252.60 & 2.72  & 0.10 &  333.8  & 0.90 \\
      & 1996.630 & 0.019 $\pm$ 0.001 & 17.731 &  250.16 & 2.72  & 0.10 &  333.8  & 1.31 \\
      & 1999.899 & 0.019 $\pm$ 0.001 & 17.641 &  248.96 & 2.72  & 0.10 &  333.8  & 1.28 \\
      & 2003.162 & 0.029 $\pm$ 0.002 & 17.763 &  249.71 & 2.72  & 0.10 &  333.8  & 2.07 \\
      & 2005.062 & 0.027 $\pm$ 0.001 & 17.861 &  249.79 & 2.72  & 0.10 &  333.8  & 1.67 \\
N1... & 1990.185 & 0.107 $\pm$ 0.024 & 17.033 &    5.15 & 13.58 & 0.27 &   20.3  & 1.13 \\ 
      & 1994.708 & 0.093 $\pm$ 0.020 & 18.678 &    4.72 & 13.58 & 0.27 &   20.3  & 0.90 \\
      & 1996.630 & 0.086 $\pm$ 0.006 & 19.143 &    5.95 & 13.58 & 0.27 &   20.3  & 1.31 \\
      & 1999.899 & 0.085 $\pm$ 0.006 & 18.874 &    5.45 & 13.58 & 0.27 &   20.3  & 1.28 \\
      & 2003.162 & 0.105 $\pm$ 0.006 & 19.271 &    6.13 & 13.58 & 0.27 &   20.3  & 2.07 \\
      & 2005.062 & 0.091 $\pm$ 0.005 & 19.353 &    6.15 & 13.58 & 0.27 &   20.3  & 1.67 \\
N2... & 1990.185 & 0.044 $\pm$ 0.010 & 21.103 &   12.86 &  2.38 & 1.00 & 254.2  & 1.13 \\ 
      & 1994.708 & 0.057 $\pm$ 0.013 & 21.414 &   11.47 &  2.38 & 1.00 & 254.2  & 0.90 \\   
      & 1996.630 & 0.049 $\pm$ 0.003 & 21.416 &   13.17 &  2.38 & 1.00 & 254.2  & 1.31 \\
      & 1999.899 & 0.056 $\pm$ 0.004 & 21.487 &   13.15 &  2.38 & 1.00 & 254.2  & 1.28 \\
      & 2003.162 & 0.068 $\pm$ 0.004 & 21.613 &   13.42 &  2.38 & 1.00 & 254.2  & 2.07 \\
      & 2005.062 & 0.051 $\pm$ 0.003 & 21.585 &   13.39 &  2.38 & 1.00 & 254.2  & 1.67 \\

\enddata
\tablenotetext{*}{NOTE - Parameters of each Gaussian component of the model
brightness distribution are as follows:  Component, Gaussian component; 
Epoch, year of observation (see Table \ref{Observations} and \S\ref{Observations});  
$S$, flux density; $r$, $\theta$, polar coordinates of the
center of the component relative to the center of component C1 (\cite{Beasley02} 
calculated the position of the source to be
$\sim$ RA 04h05m49.2623s Dec +38$^o$03$^{\prime}$32$^{\prime\prime}$.235. We are assuming component C1 to be approximately
at this location)
; $a$, semimajor axis; $b/a$, axial ratio; 
$\Phi$, component orientation; $\chi^{2}$, goodness-of-fit for eight component model in each epoch. 
All angles are measured
from north through east.  Errors in flux are based on our absolute amplitude calibration
as well as the rms noise. Note that due to the complicated morphology of the source, variability
studies were performed using components C1, C2, the sum of the southern components (S1, S2,
S3, and S4), and the sum of the northern components (N1 and N2).}

\end{deluxetable}

\begin{deluxetable}{lccccccc}
\tabletypesize{\scriptsize}
\tablecolumns{8}
\tablewidth{0pt}
\tablecaption{Component Motion Fitting Results.\label{Motions_Table}}
\tablehead{\colhead{Component}&\colhead{Velocity in $y$}&\colhead{$\chi^{2}$}&\colhead{Velocity in $x$}&\colhead{$\chi^{2}$}&\colhead{Velocity}&\colhead{Velocity}&\colhead{Angle of Motion}\tablenotemark{*} \\
\colhead{} & \colhead{(mas/y)} & \colhead{}&\colhead{(mas/y)}&\colhead{}&\colhead{(mas/y)}&\colhead{(c)}&\colhead{($^o$)}}
\startdata
C1 & Reference Component &... &... &... &... &... &...  \\
C2 & $-$0.006 $\pm$ 0.007 & 2.02 & \phs0.003 $\pm$ 0.007 & 0.60 & 0.007 $\pm$ 0.009 & 0.023 $\pm$ 0.032 & 153.4  \\
S1 & $-$0.027 $\pm$ 0.003 & 9.66 & $-$0.024 $\pm$ 0.003 & 6.90 & 0.036 $\pm$ 0.004 & 0.124 $\pm$ 0.015 & 221.6  \\
S2 & \phs0.007 $\pm$ 0.002 & 0.35 & $-$0.002 $\pm$ 0.002 & 1.34 & 0.007 $\pm$ 0.002 & 0.025 $\pm$ 0.009  & 344.5  \\
S3 & $-$0.013 $\pm$ 0.002 & 1.30 & \phs0.010 $\pm$ 0.002 & 0.34 & 0.016 $\pm$ 0.003 & 0.056 $\pm$ 0.010 & 142.4  \\
S4 & $-$0.016 $\pm$ 0.006 & 5.85 & $-$0.026 $\pm$ 0.006 & 1.04 & 0.031 $\pm$ 0.008 & 0.105 $\pm$ 0.028 & 238.4  \\
N1 & \phs0.046 $\pm$ 0.006 & 1.90 & \phs0.028 $\pm$ 0.006 & 1.30 & 0.054 $\pm$ 0.008 & 0.185 $\pm$ 0.028 & 31.3  \\
N2 & \phs0.023 $\pm$ 0.004 & 0.68 & \phs0.024 $\pm$ 0.004 & 2.59 & 0.033 $\pm$ 0.006 & 0.114 $\pm$ 0.019 & 46.2  \\
\enddata
\tablenotetext{*}{Angles measured from north through east.}
\end{deluxetable}

\begin{deluxetable}{lccccccc}
\tabletypesize{\scriptsize}
\tablecolumns{8}
\tablewidth{0pt}
\tablecaption{Continuum Spectrum Results\tablenotemark{*}.\label{Peak_Fluxes}}
\tablehead{\colhead{Component}&\colhead{Frequency}&\colhead{Flux}
&\colhead{$\alpha_{5-8}$}&\colhead{$\alpha_{8-15}$}&\colhead{$\alpha_{8-22}$}&\colhead{$\alpha_{15-22}$}
&\colhead{$\alpha_{22-43}$}\\
\colhead{} &\colhead{(GHz)}&\colhead{(mJ)} &\colhead{} 
&\colhead{}&\colhead{}&\colhead{}&\colhead{}}
\startdata
C1... &  4.987 & 53.2 $\pm$ 2.6 &\phs0.25 $\pm$ 0.19&$-$0.24 $\pm$ 0.25&$-$0.55 $\pm$ 0.15 &$-$1.07 $\pm$ 0.54& $-$0.62 $\pm$ 0.30\\
      &  8.421 & 60.5 $\pm$ 3.0 &               &                & &               &                   \\
      &  15.365& 52.4 $\pm$ 5.2 &               &                & &               &                  \\
      &  22.233& 35.3 $\pm$ 3.5 &               &                & &               &                  \\
      &  43.217& 23.4 $\pm$ 2.3 &               &                & &               &                  \\
C2... &  4.987 & 10.7 $\pm$ 0.5 &\phs0.68 $\pm$ 0.19&$-$0.06 $\pm$ 0.26&$-$0.37 $\pm$ 0.16 &$-$0.88 $\pm$ 0.55&$-$0.44 $\pm$ 0.31  \\
      &  8.421 & 15.3 $\pm$ 0.8 &               &                & &               &                  \\
      &  15.365& 14.8 $\pm$ 1.5 &               &                & &               &                  \\
      &  22.233& 10.7 $\pm$ 1.1 &               &                & &               &                  \\
      &  43.217&  8.0 $\pm$ 0.8 &               &                & &               &                  \\
S2... &  4.987 & 72.2 $\pm$ 3.5 &$-$0.42 $\pm$ 0.19&$-$0.73 $\pm$ 0.25&$-$1.11 $\pm$ 0.16 &$-$1.73 $\pm$ 0.55&$-$1.24 $\pm$ 0.31\\
      &  8.421 & 57.8 $\pm$ 3.0 &                &                & &               &                  \\
      &  15.365& 37.2 $\pm$ 3.7 &                &                & &               &                  \\
      &  22.233& 19.6 $\pm$ 2.0 &                &                & &               &                  \\
      &  43.217&  8.6 $\pm$ 0.9 &                &                & &               &                  \\
N2... &  4.987 & 11.2 $\pm$ 0.5 &$-$0.18 $\pm$ 0.18&$-$0.56 $\pm$ 0.24&$-$0.89 $\pm$ 0.15 &$-$1.43 $\pm$ 0.51&      ...        \\
      &  8.421 & 10.2 $\pm$ 0.5 &                &                & &               &                  \\
      &  15.365&  7.3 $\pm$ 0.7 &                &                & &               &                  \\
      &  22.233&  4.3 $\pm$ 0.4 &                &                & &               &                  \\
      &  43.217& undetected     &                &                & &               &                  \\
\enddata
\tablenotetext{*}{Results obtained from the 2005 VLBA observations. 
The fluxes were measured from matching resolution images.
See \S\ref{RadioContSpec} for more details.}
\end{deluxetable}

\clearpage

\setcounter{figure}{5}
\begin{figure}
\centering
\scalebox{0.5}{\includegraphics{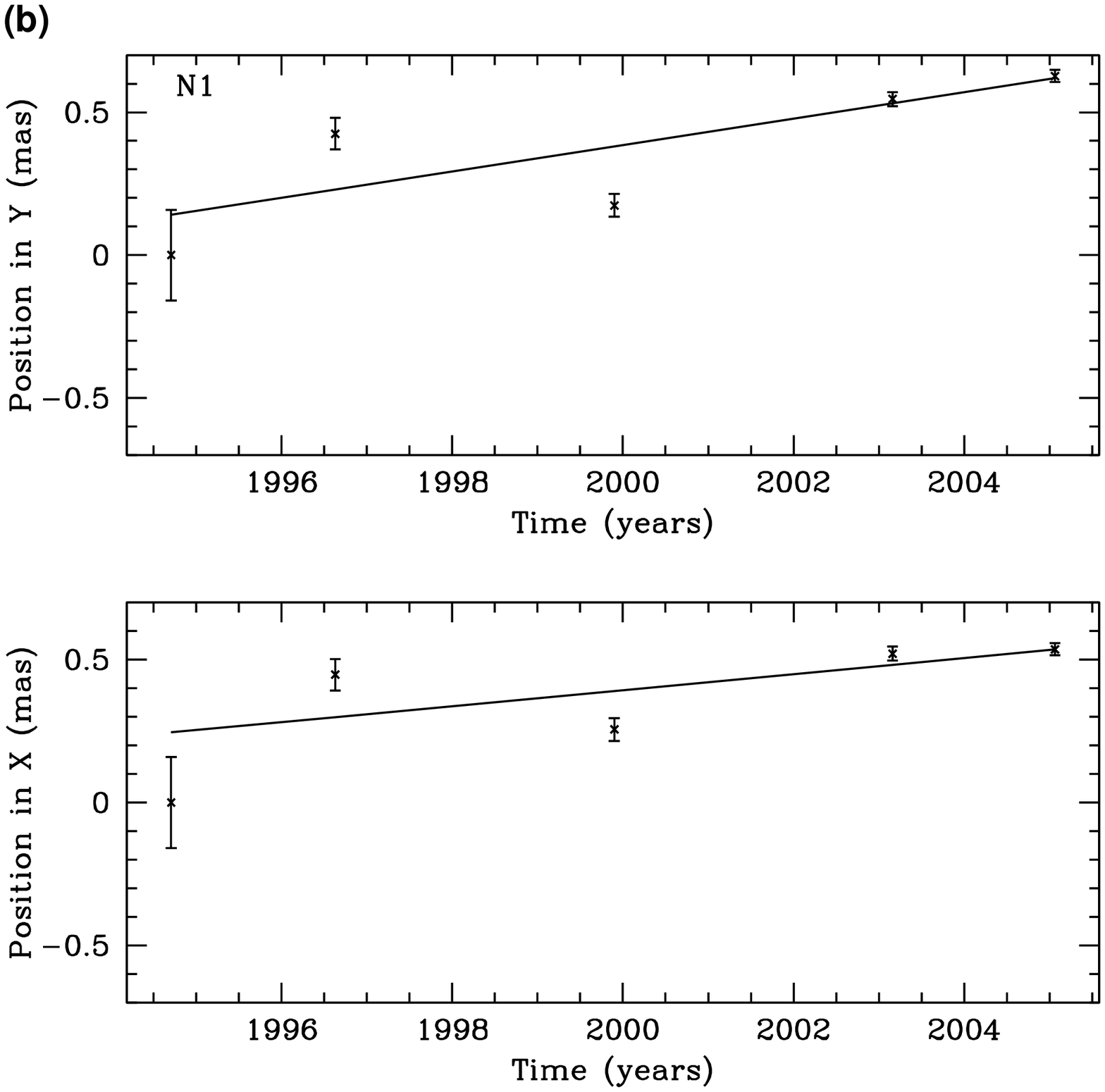}}
\vspace{10mm}
\caption{Figure \ref{C2} continued}
\label {N1}
\end{figure}

\setcounter{figure}{5}
\begin{figure}
\centering
\scalebox{0.5}{\includegraphics{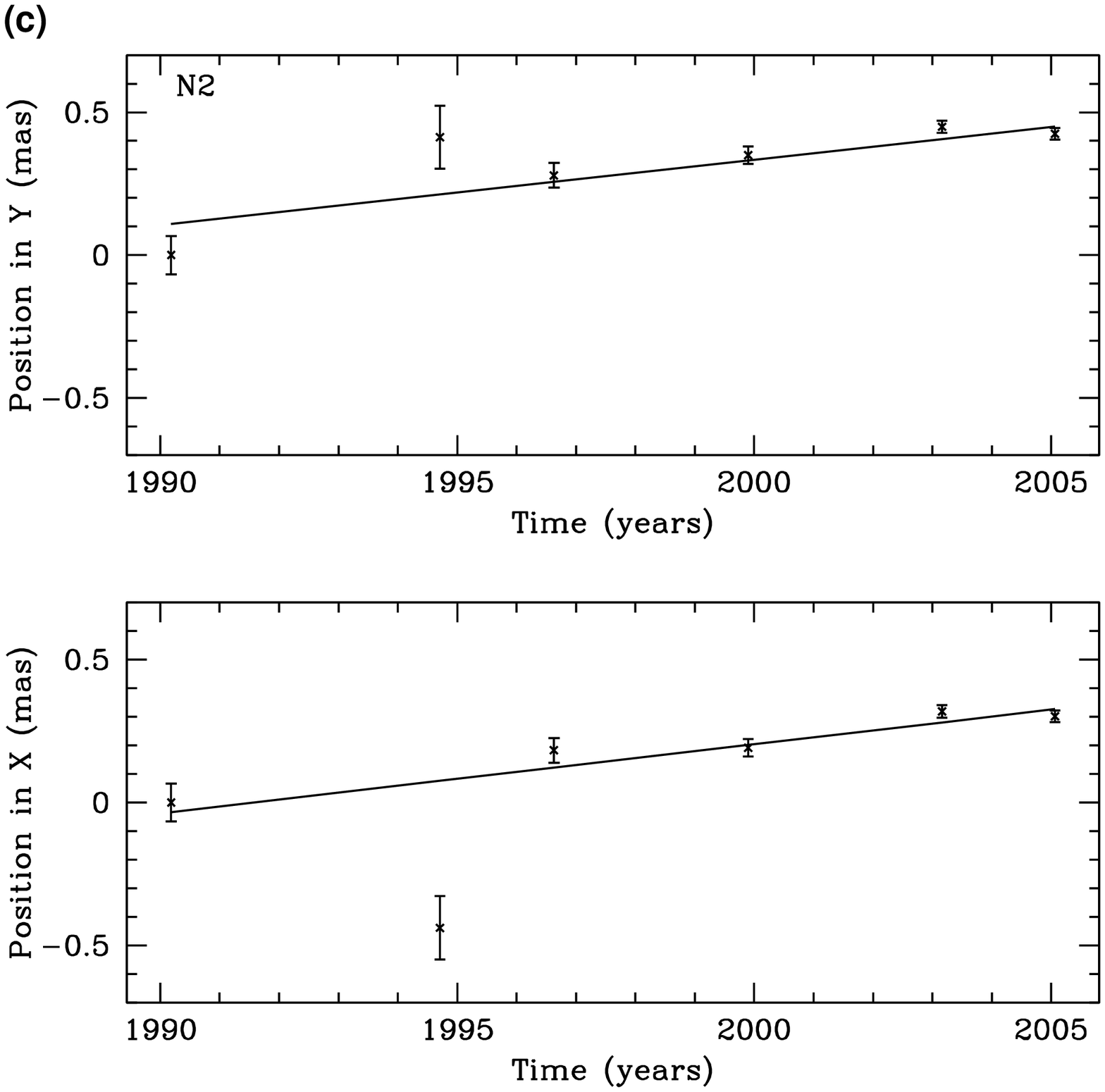}}
\vspace{10mm}
\caption{Figure \ref{C2} continued}
\label{N2}
\end{figure}

\setcounter{figure}{5}
\begin{figure}
\centering
\scalebox{0.5}{\includegraphics{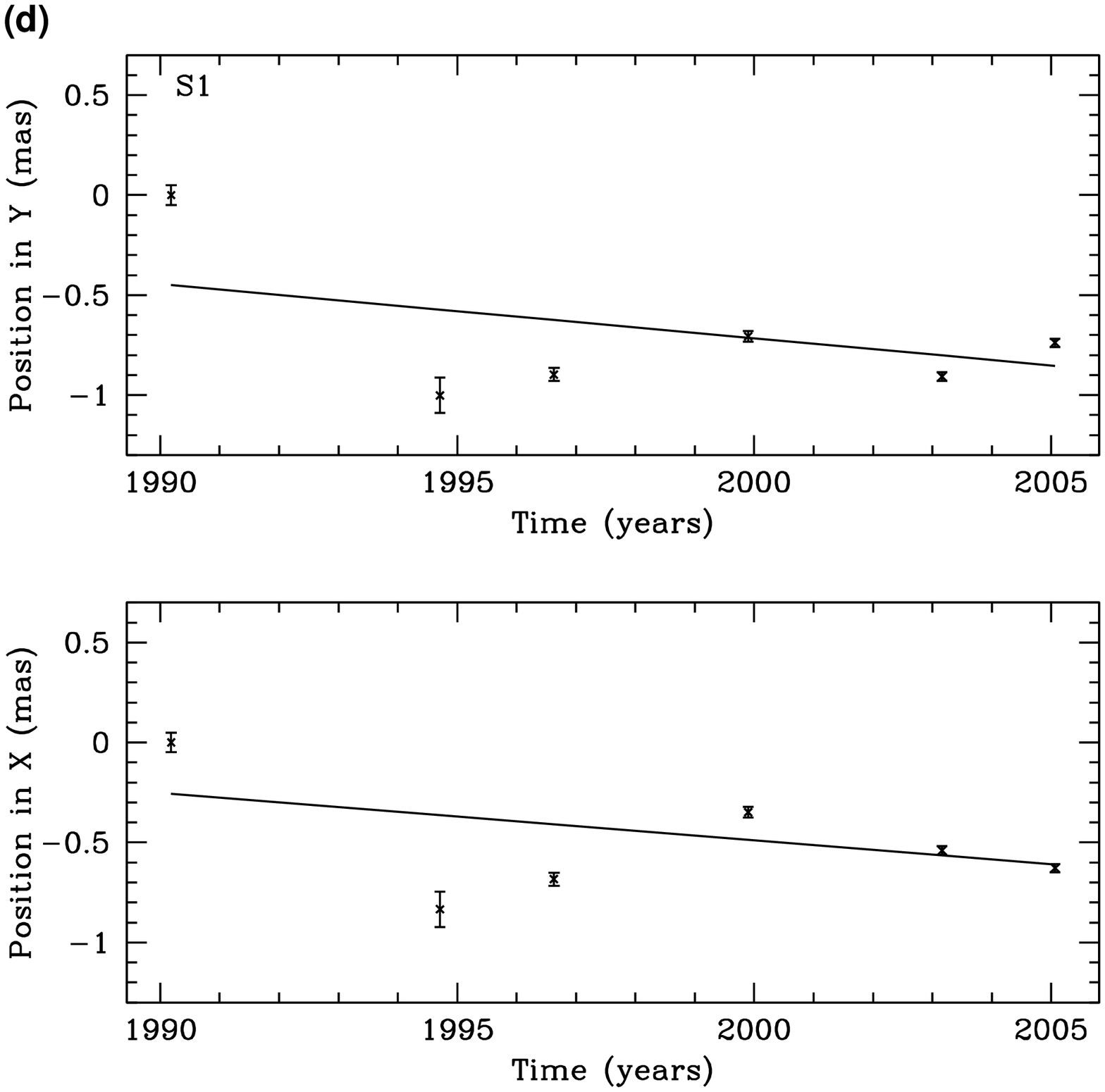}}
\vspace{10mm}
\caption{Figure \ref{C2} continued}
\label{S1}
\end{figure}

\setcounter{figure}{5}
\begin{figure}
\centering
\scalebox{0.5}{\includegraphics{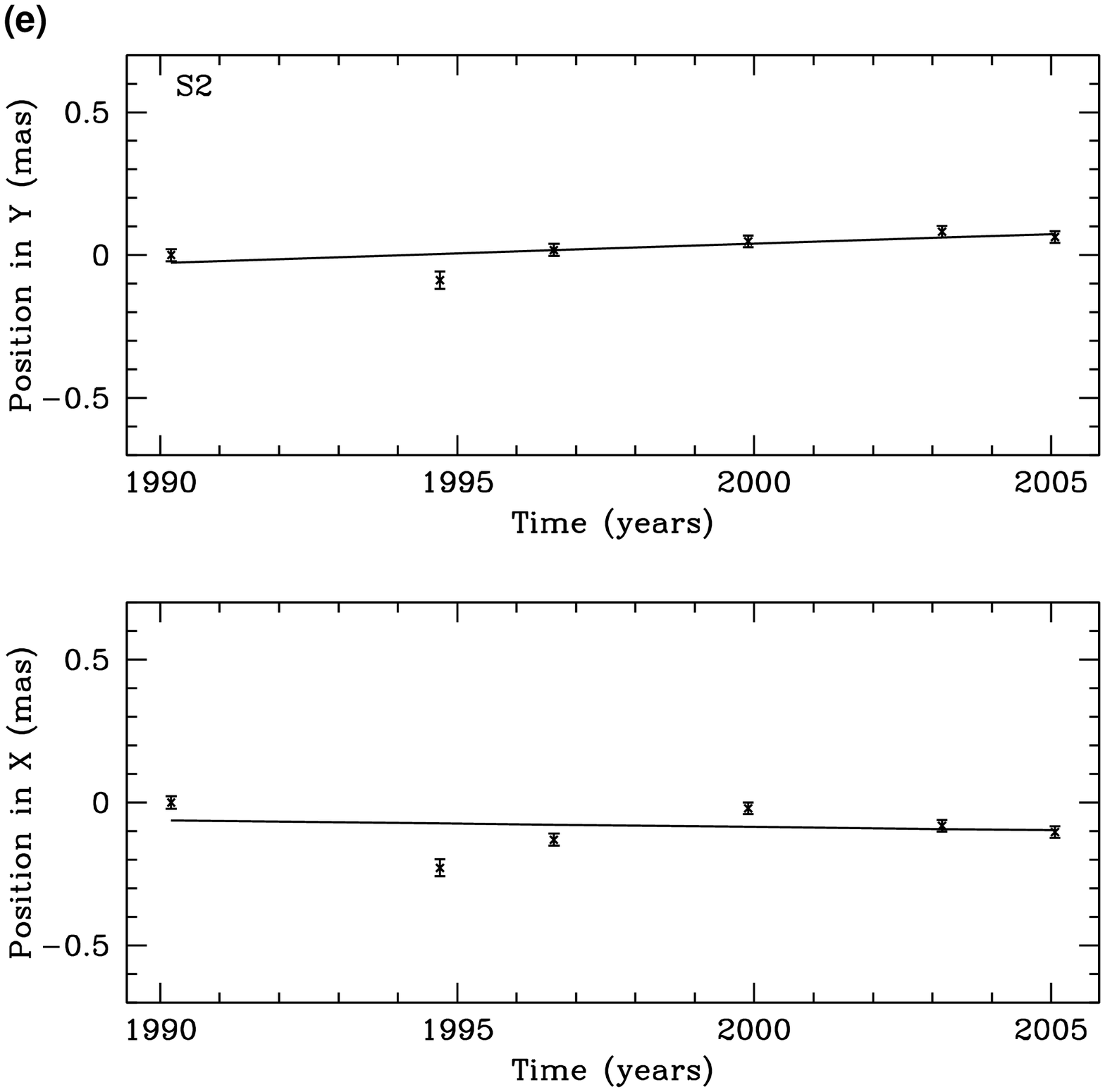}}
\vspace{10mm}
\caption{Figure \ref{C2} continued}
\label{S2}
\end{figure}

\setcounter{figure}{5}
\begin{figure}
\centering
\scalebox{0.5}{\includegraphics{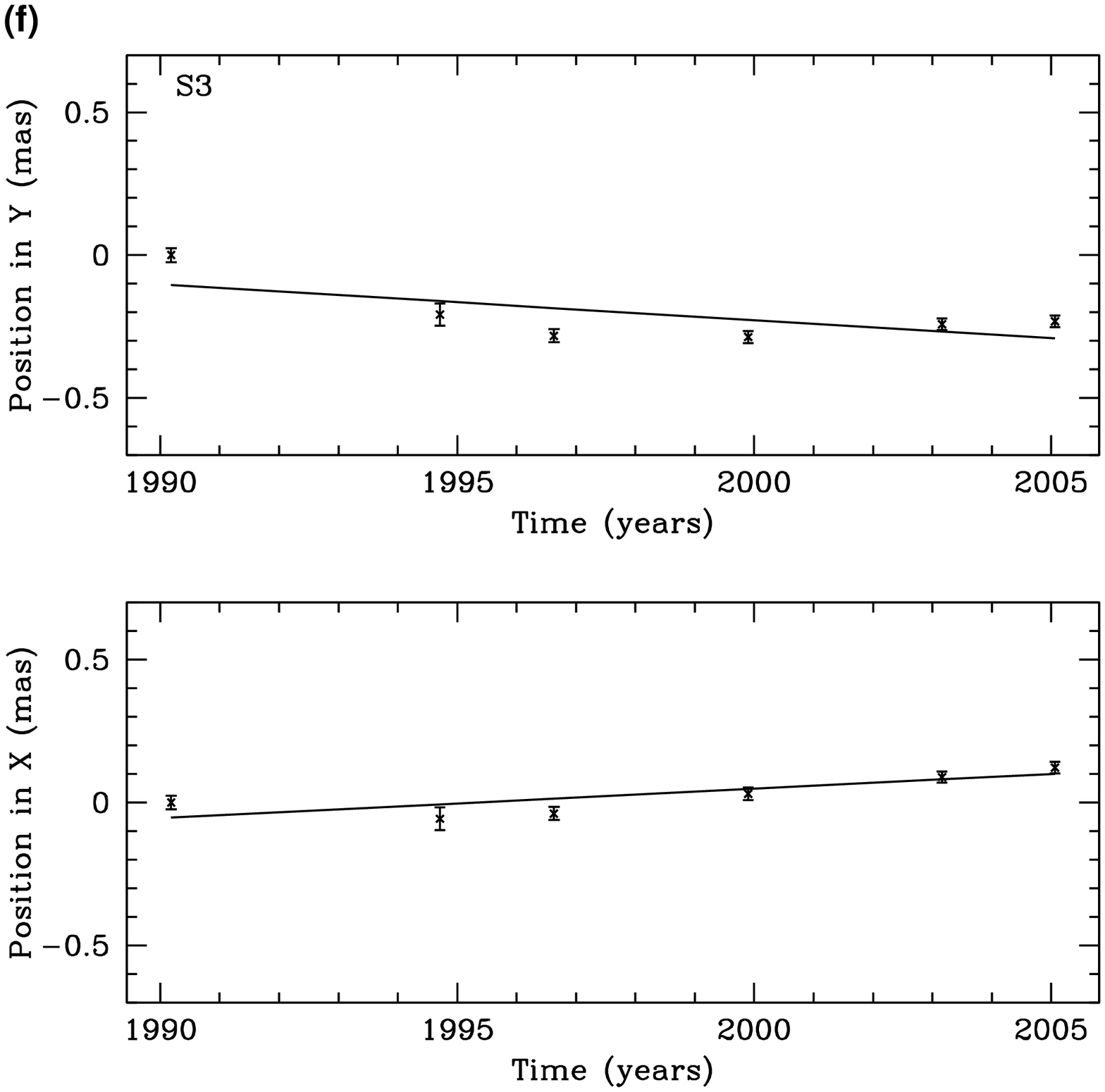}}
\vspace{10mm}
\caption{Figure \ref{C2} continued}
\label{S3}
\end{figure}

\setcounter{figure}{5}
\begin{figure}
\centering
\scalebox{0.5}{\includegraphics{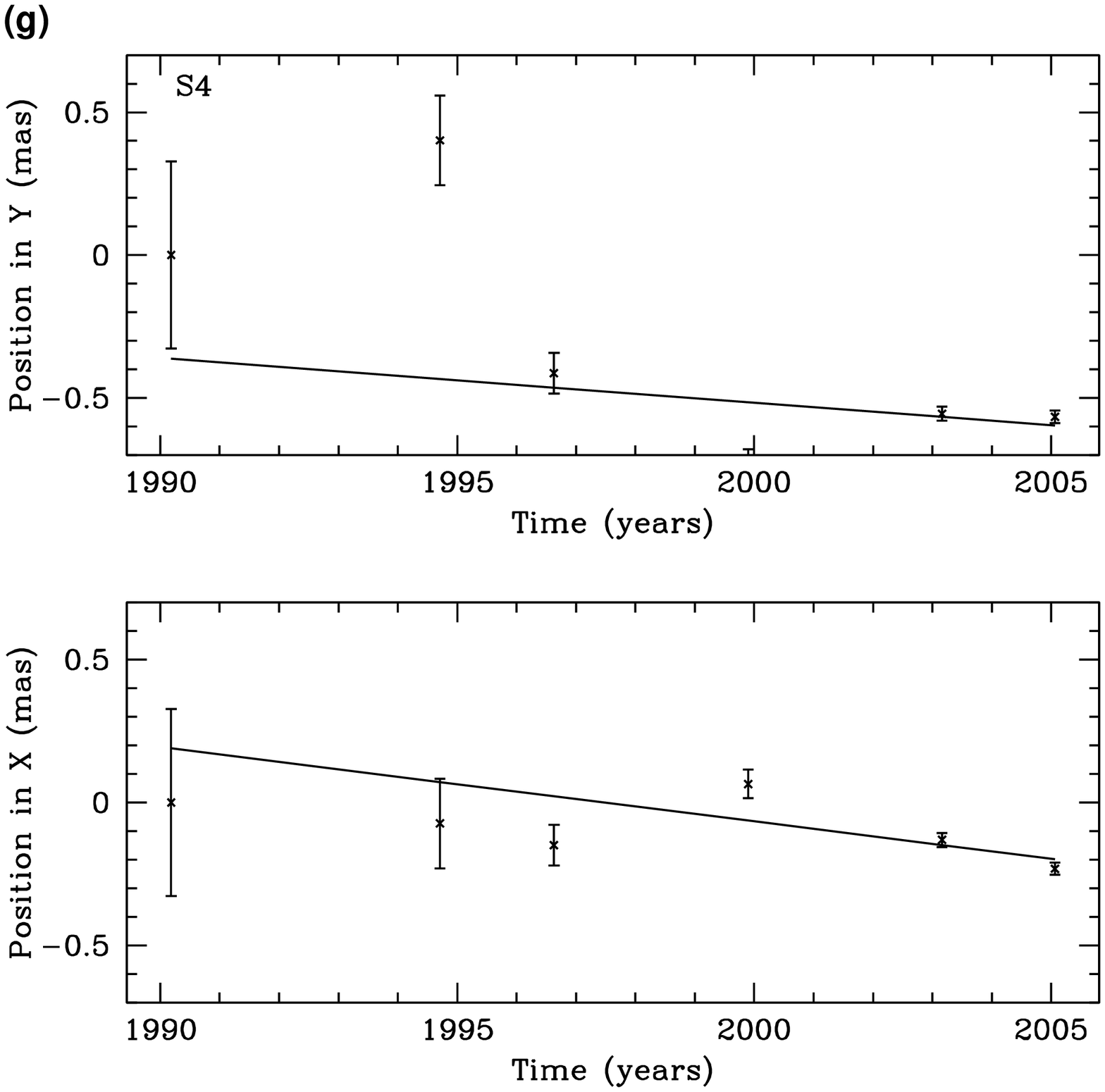}}
\vspace{10mm}
\caption{Figure \ref{C2} continued}
\label{S4}
\end{figure}

\begin{figure}
\centering
\scalebox{1.00}{\includegraphics{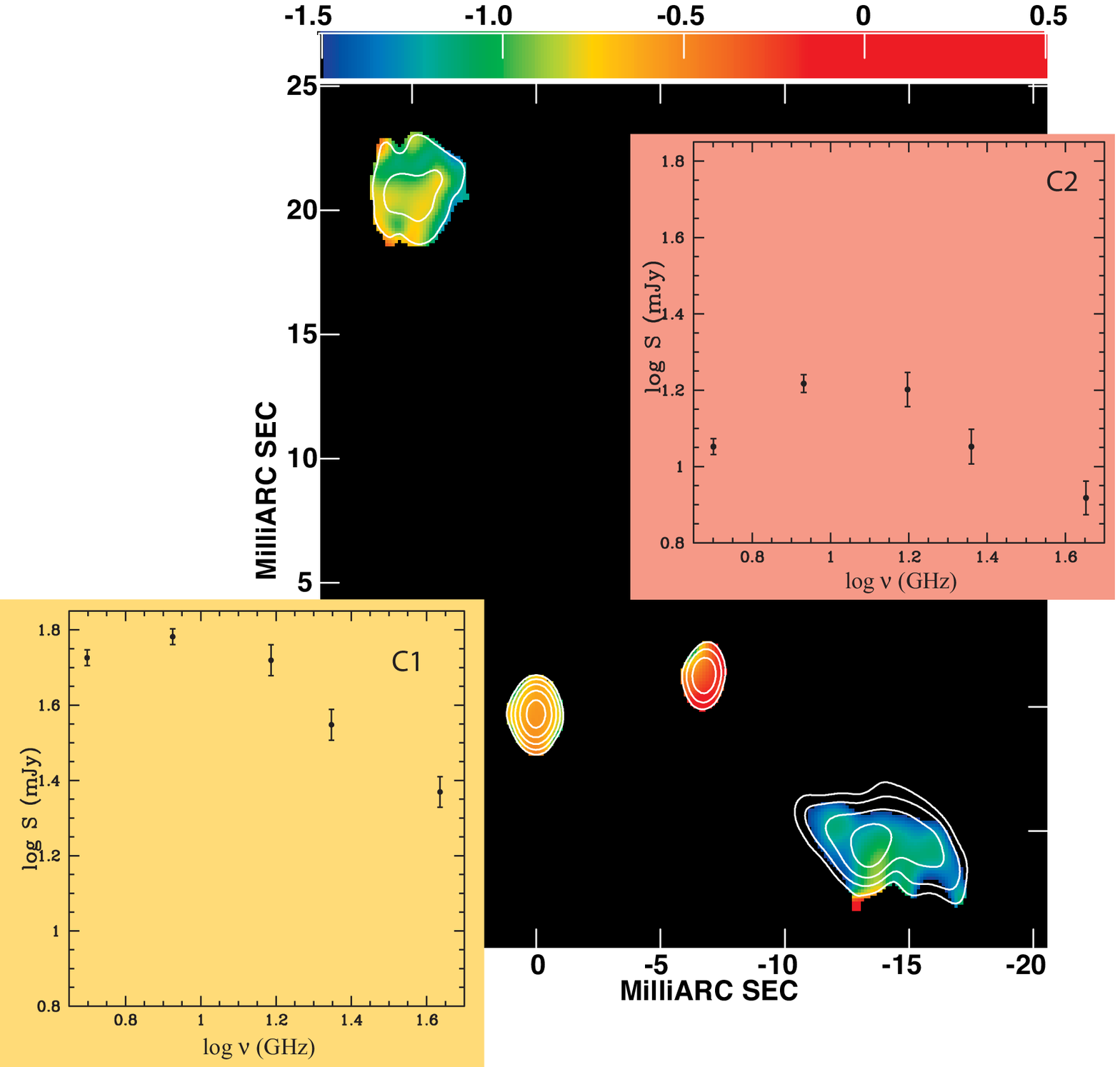}}
\vspace{1cm}
\caption{Spectral index distribution between 8 and 22 GHz from the 2005 VLBA observations.
The contours are taken from the 22 GHz observations and are set at 7$\sigma$, increasing 
by a factor of 2 thereafter.}
\label{VLBA05spix8-22}
\end{figure}


\begin{thebibliography}{}


\bibitem[Barkhouse \& Hall(2001)]{Bark01} Barkhouse, W.~A., \& 
Hall, P.~B.\ 2001, \aj, 121, 2843 


\bibitem[Beasley et al.(2002)]{Beasley02} Beasley, A.~J., Gordon, 
D., Peck, A.~B., Petrov, L., MacMillan, D.~S., Fomalont, E.~B., \& Ma, C.\ 
2002, \apjs, 141, 13

\bibitem[Becker et al.(1991)]{bwe} Becker, R.~H., White, 
R.~L., \& Edwards, A.~L.\ 1991, \apjs, 75, 1   

\bibitem[Begelman et al.(1980)]{Begelman80} Begelman, M.~C., 
Blandford, R.~D., \& Rees, M.~J.\ 1980, \nat, 287, 307 

\bibitem[Brinkmann et al.(2000)]{brink} Brinkmann, W., 
Laurent-Muehleisen, S.~A., Voges, W., Siebert, J., Becker, R.~H., 
Brotherton, M.~S., White, R.~L., \& Gregg, M.~D.\ 2000, \aap, 356, 445 


\bibitem[Britzen et al.(2003)]{bri03} Britzen, S., et al.\ 
2003, Future Directions in High Resolution Astronomy: A Celebration of the 
10th Anniversary of the VLBA, edited by J.~D.~Romney and 
M.~J.~Reid.~Socorro, N.M.~: National Radio Astronomy Observatory, 2003., 
p.85, 85 


\bibitem[Cutri et al.(2003)]{2mass} Cutri, R. et al. 2003, 2MASS 
Extended Source Catalog, NASA/IPAC Extragalactic Database

\bibitem[Filippenko \& Ho(2003)]{Filipenko03} Filippenko, A.~V., \& 
Ho, L.~C.\ 2003, \apjl, 588, L13 

\bibitem[Gebhardt et al.(2005)]{Gebhardt05} Gebhardt, K., Rich, 
R.~M., \& Ho, L.~C.\ 2005, \apj, 634, 1093 

\bibitem[Hill et al.(1998)]{Hill98} Hill, G.~J., Nicklas, 
H.~E., MacQueen, P.~J., Tejada, C., Cobos Duenas, F.~J., \& Mitsch, W.\ 
1998, \procspie, 3355, 375 
 

\bibitem[Hughes(2003)]{hughes} Hughes, S.~A.\ 2003, Annals of 
Physics, 303, 142

\bibitem[Komossa(2003a)]{Komossa03} Komossa, S.\ 2003, AIP 
Conf.~Proc.~686: The Astrophysics of Gravitational Wave Sources, 686, 161 

\bibitem[Komossa et al.(2003b)]{Komo03} Komossa, S., Burwitz, 
V., Hasinger, G., Predehl, P., Kaastra, J.~S., \& Ikebe, Y.\ 2003, \apjl, 
582, L15 

\bibitem[Kormendy \& Gebhardt(2001)]{Kormendy01} Kormendy, J., \& 
Gebhardt, K.\ 2001, AIP Conf.~Proc.~586: 20th Texas Symposium on 
relativistic astrophysics, 586, 363 

\bibitem[Lister(2001)]{lis01} Lister, M.~L.\ 2001, \apj, 562, 
208 

\bibitem[Lind \& Blandford(1985)]{Lind&Blandford85} Lind, K.~R., \& 
Blandford, R.~D.\ 1985, \apj, 295, 358

\bibitem[Maness et al.(2004)]{man03} Maness, H.~L., Taylor, 
G.~B., Zavala, R.~T., Peck, A.~B., \& Pollack, L.~K.\ 2004, \apj, 602, 123 

\bibitem[Merritt \& Milosavljevi{\'c}(2005)]{Milo05} Merritt, 
D., \& Milosavljevi{\'c}, M.\ 2005, Living Reviews in Relativity, 8, 8 

\bibitem[Owen et al.(1985)]{Owen85} Owen, F.~N., Odea, C.~P., 
Inoue, M., \& Eilek, J.~A.\ 1985, \apjl, 294, L85 
 

\bibitem[Pauliny-Toth et al.(1978)]{pt} Pauliny-Toth, 
I.~I.~K., Witzel, A., Preuss, E., K{$\rm \ddot{u}$}hr, H., Fomalont, E.~B., Davis, M.~M., 
\& Kellermann, K.~I.\ 1978, \aj, 83, 451

\bibitem[Peters(1964)]{Peters64} Peters, P. C.\ 1964, \prb, 136, 1224

\bibitem[Ramsey et al.(1998)]{Ramsey98} Ramsey, L.~W., et al.\ 
1998, \procspie, 3352, 34 


\bibitem[Richstone et al.(1998)]{Rich98} Richstone, D., et 
al.\ 1998, \nat, 395, A14 

\bibitem[Sesana et al.(2004)]{Sesana04} Sesana, A., Haardt, F., 
Madau, P., \& Volonteri, M.\ 2004, \apj, 611, 623 

\bibitem[Sesana et al.(2005)]{Sesana05} Sesana, A., Haardt, F., 
Madau, P., \& Volonteri, M.\ 2005, \apj, 623, 23 
 
\bibitem[Shepherd et al.(1995)]{Shep95} Shepherd, M. C., 
Pearson, T.J., \& Taylor, G.B.\ 1995, BASS, 27, 903

\bibitem[Sillanp{\" a}{\" a} et al.(1988)]{Sillanpaa88} Sillanp{\" a}{\" a}, A., 
Haarala, S., Valtonen, M.~J., Sundelius, B., \& Byrd, G.~G.\ 1988, \apj, 
325, 628 

\bibitem[Stickel et al.(1993)]{Stickel93} Stickel, M., 
Kuehr, H., \& Fried, J.~W.\ 1993, \aaps, 97, 483 

\bibitem[Taylor et al.(1996)]{Taylor96} Taylor, 
G.~B., Readhead, A.~C.~S., \& Pearson, T.~J.\ 1996, \apj, 463, 95 

\bibitem[Taylor \& Vermeulen(1997)]{Taylor&Vermeulen97} Taylor, G.~B., \& 
Vermeulen, R.~C.\ 1997, \apjl, 485, L9

\bibitem[Taylor et al.(2005)]{Taylor05} Taylor, G.~B., et al.\ 
2005, \apjs, 159, 27 

\bibitem[Ulvestad et al.(2001)]{Ulvestad01} Ulvestad, J.,
Greisen, E.~W. \& Mioduszewski, A.\ 2001, AIPS Memo 105:AIPS Procedures
for initial VLBA Data Reduction, NRAO

\bibitem[Valtaoja et al.(2000)]{Valtaoja00} Valtaoja, E., 
Ter{\"a}sranta, H., Tornikoski, M., Sillanp{\"a}{\"a}, A., Aller, M.~F., 
Aller, H.~D., \& Hughes, P.~A.\ 2000, \apj, 531, 744 


\bibitem[van Moorsel et al.(1996)]{vanMoorsel96} van 
Moorsel, G., Kemball, A., \& Greisen, E.\ 1996, ASP Conf.~Ser.~101: 
Astronomical Data Analysis Software and Systems V, 5, 37 



\bibitem[Voges et al.(1999)]{rosat} Voges, W., et al.\ 1999, 
\aap, 349, 389


\bibitem[Wills et al.(1973)]{Wills73} Wills, B.~J., Wills, D., 
\& Douglas, J.~N.\ 1973, \aj, 78, 521 


\bibitem[Xu et al.(1995)]{Xu95} Xu, W., Readhead, A.~C.~S., 
Pearson, T.~J., Polatidis, A.~G., \& Wilkinson, P.~N.\ 1995, \apjs, 99, 297 

 
\end{thebibliography}
\end{document}